\documentclass[aps,prx,lengthcheck,superscriptaddress,reprint,showpacs,longbibliography,nofootinbib,amsmath,amssymb]{revtex4-1}
\usepackage[utf8]{inputenc}
\usepackage[T1]{fontenc}

\usepackage{amssymb}
\usepackage{graphicx}
\usepackage[usenames,dvipsnames]{xcolor}
\usepackage[colorlinks,linkcolor=blue,citecolor=blue,urlcolor=blue]{hyperref}

\def\equationautorefname~#1\null{Eq.~(#1)\null}

\renewcommand{\appendixautorefname}{appendix}
\newcommand{\appref}[1]{\href{#1}{\appendixautorefname~\ref*{#1}}}

\newcommand{\vecop}[1]{\hat{\textbf{#1}}}
\newcommand{\vecopsym}[1]{\hat{\boldsymbol{#1}}}
\newcommand{\blambda}{\boldsymbol{\lambda}}
\newcommand{\bmu}{\boldsymbol{\mu}}
\newcommand{\bR}{\textbf{R}}
\newcommand{\br}{\textbf{r}}

\newcommand{\bE}{\textbf{E}}

\begin{document}

\title{Cavity Casimir-Polder forces and their effects in ground state chemical reactivity}

\author{Javier Galego}
\email{javier.galego@uam.es}
\affiliation{Departamento de F{\'\i}sica Te{\'o}rica de la Materia Condensada and Condensed Matter Physics Center (IFIMAC), Universidad Aut\'onoma de Madrid, E-28049 Madrid, Spain}
\author{Cl\`audia Climent}
\email{claudia.climent@uam.es}
\affiliation{Departamento de F{\'\i}sica Te{\'o}rica de la Materia Condensada and Condensed Matter Physics Center (IFIMAC), Universidad Aut\'onoma de Madrid, E-28049 Madrid, Spain}
\author{Francisco~J.~Garcia-Vidal}
\email{fj.garcia@uam.es}
\affiliation{Departamento de F{\'\i}sica Te{\'o}rica de la Materia Condensada and Condensed Matter Physics Center (IFIMAC), Universidad Aut\'onoma de Madrid, E-28049 Madrid, Spain}
\affiliation{Donostia International Physics Center (DIPC), E-20018 Donostia/San Sebastian, Spain}
\author{Johannes Feist}
\email{johannes.feist@uam.es}
\affiliation{Departamento de F{\'\i}sica Te{\'o}rica de la Materia Condensada and Condensed Matter Physics Center (IFIMAC), Universidad Aut\'onoma de Madrid, E-28049 Madrid, Spain}

\begin{abstract}
Here we present a fundamental study on how the ground-state chemical reactivity
of a molecule can be modified in a QED scenario, i.e., when it is placed inside
a cavity and there is strong coupling between the cavity field and vibrational
modes within the molecule. We work with a model system for the molecule
(Shin--Metiu model) in which nuclear, electronic and photonic degrees of freedom
are treated on the same footing. This simplified model allows the comparison of
exact quantum reaction rate calculations with predictions emerging from
transition state theory based on the cavity Born--Oppenheimer approach. We
demonstrate that QED effects are indeed able to significantly modify activation
barriers in chemical reactions and, as a consequence, reaction rates. The
critical physical parameter controlling this effect is the permanent dipole of
the molecule and how this magnitude changes along the reaction coordinate. We
show that the effective coupling can lead to significant single-molecule energy
shifts in an experimentally available nanoparticle-on-mirror cavity. We then
apply the validated theory to a realistic case (internal rotation in the
1,2-dichloroethane molecule), showing how reactions can be inhibited or
catalyzed depending on the profile of the molecular dipole. Furthermore, we
discuss the absence of resonance effects in this process, which can be
understood through its connection to Casimir--Polder forces. Finally, we treat
the case of many-molecule strong coupling, and find collective modifications of
reaction rates if the molecular permanent dipole moments are oriented with
respected to the cavity field. This demonstrates that collective coupling can
also provide a mechanism for modifying ground-state chemical reactivity of an
ensemble of molecules coupled to a cavity mode.
\end{abstract}

\maketitle

\section{Introduction}


The field of (non-relativistic) cavity quantum electrodynamics (CQED) has proved
that the quantum nature of light can be exploited for many interesting
applications that involve the modifications of material properties in one way or
another~\cite{Miller2005,Walther2006}. In this context, strong light-matter
coupling is particularly appealing~\cite{Thompson1992}. The regime of strong
coupling is achieved when the coherent energy exchange between the excitations
of a material (excitons) and of the cavity light modes is faster than the decay
rate of either constituent. The resulting excitations are the well-known
polaritons, which combine properties of both light and matter, leading to many
interesting applications (see~\cite{Sanvitto2016} for a recent review). In
recent years, strong coupling to organic materials has received great attention
for its potential to greatly influence fundamental features of the underlying
organic molecules such as their optical response~\cite{Barachati2015, Cwik2016,
Zeb2018}, transport properties~\cite{Orgiu2015, Feist2015, Zhong2017,
Garcia-Vidal2017, Saez-Blazquez2018Organic}, or chemical
reactivity~\cite{Hutchison2012, Thomas2016, Munkhbat2018}. In particular, the
potential of polaritonic chemistry, i.e., the ability to influence the chemical
structure and reactions of organic compounds through coupling to a cavity, has
attracted a lot of interest~\cite{Tokatly2013, Galego2015, Flick2015,
Galego2016, Herrera2016, Kowalewski2016Cavity, Kowalewski2016Nonadiabatic,
Galego2017, Martinez-Martinez2018Polariton, Flick2017Atoms, Flick2017Cavity,
Luk2017, Feist2018, Keeling2018, Martinez-Martinez2018Can, Ruggenthaler2018,
Ribeiro2018, Vendrell2018Coherent, Vendrell2018Collective}.

Most of the research on polaritonic chemistry with organic molecules has dealt
with electronic strong coupling. This leads to many interesting effects such as
collective protection of polaritons and changes in chemical
reactivity~\cite{Galego2016,Herrera2016}, cavity-induced nonadiabatic
phenomena~\cite{Kowalewski2016Cavity, Feist2018, Vendrell2018Collective}, and
the opening of novel reaction pathways in photochemistry~\cite{Galego2017}.

More recently, the possibility of influencing the thermally driven reactivity of
organic molecules in the \emph{ground} state has been demonstrated by coupling
the cavity to vibrational transitions of the molecules~\cite{Thomas2016,
Hiura2018, Lather2018, Thomas2019}. This opens a wide range of possibilities due
to the fact that no external input of energy is needed at all, such as
cavity-enabled catalysis and manipulation of ground-state chemical reactions.
Cavity-induced modifications to the ground state have also been theoretically
studied. In particular, for model molecules without ground-state dipole moments
and only electronic dipole transitions, it has been shown that there is no
collective enhancement of energy shifts~\cite{Galego2015}, and more
specifically, that chemical reactions are not strongly modified even under
ultrastrong collective coupling~\cite{Martinez-Martinez2018Can}. In a series of
papers based on more microscopic models, Flick and co-workers have shown that
ground state properties can be significantly modified under single-molecule
(ultra-)strong coupling~\cite{Flick2015, Flick2017Atoms, Flick2017Cavity}, but
have not treated chemical reactivity.


In the present work, we aim to understand cavity-induced modifications of
ground-state chemistry in coupled molecule-cavity systems. It is structured as
follows: In \autoref{sec:theory} we present the light-matter interaction
Hamiltonian for a single molecule coupled to a nanoscale cavity. After a brief
discussion of the validity of this Hamiltonian, we study a simple model system,
the Shin--Metiu model and in \autoref{sec:rates} obtain the cavity-modified
reactivity from formally exact quantum rate calculations~\cite{Yamamoto1960,
Miller1974, Miller1983}. In \autoref{sec:CBOA}, we develop a simplified theory
that allows to understand ground-state chemical reactivity changes based on
well-known concepts such as transition state theory
(TST)~\cite{Eyring1935,Laidler1987} by exploiting the cavity Born--Oppenheimer
approximation~\cite{Flick2017Atoms}. We show in \autoref{sec:pert_th} that, to a
good approximation, perturbation theory can be used to predict cavity-induced
chemical changes in terms of bare-molecule ground-state properties, and also
allows to make explicit connections to electrostatic, van der Waals, and
Casimir--Polder interactions. This is exploited in \autoref{sec:NPoM} to
demonstrate that for a realistic experimental geometry, a multimode
nanoparticle-on-mirror cavity~\cite{Chikkaraddy2016,Benz2016,Urbieta2018}, the
effective single-molecule coupling can be significant. In \autoref{sec:dich}, we
study the modification of reaction rates in the 1,2-dichloroethane molecule,
demonstrating the potential of a cavity to catalyze or inhibit reactions, or
even to modify the equilibrium configuration of the molecule. In
\autoref{sec:resonance}, we discuss in detail the dependence of chemical
reaction rates on the frequency of the cavity mode. We observe that, in contrast
to polariton formation, which requires the cavity photon and molecular
excitations to be resonant, no such requirement exists for the change of
reaction rates in the cavity. In the case of a single molecule as treated up to
that point, the coupling strengths required to obtain significant changes in
chemical reactivity correspond to the most tightly confined plasmonic nanogap
cavities available
experimentally~\cite{Kim2015,Chikkaraddy2016,Benz2016,Urbieta2018}. In
\autoref{sec:collective}, we thus extend our model to an ensemble of molecules
and find a collective enhancement of the effect under orientational alignment of
the molecular dipoles. 

We mention here that we do not explicitly treat the case of many molecules
coupled to a cavity with a continuum of modes, i.e., the case which corresponds
to the experimentally used Fabry-Perot cavities with in-plane
dispersion~\cite{Thomas2016, Thomas2019}. For the sake of simplicity, we also
neglect solvent effects. While these are well-known to be important in chemical
reactions, their effect depends strongly on the chosen solvent and experimental
setup (particularly in nanocavities). However, we mention that the latest
experimental studies indicate that solvent effects might be responsible and/or
relevant for the experimentally observed resonance-dependent ground-state
chemical reactivity~\cite{Hiura2018,Lather2018}.

\section{Theory and model system}\label{sec:theory}


\subsection{Light-matter Hamiltonian}

We start from the general non-relativistic light-matter Hamiltonian of QED in
minimal coupling, describing a collection of charged particles coupled to the
electromagnetic (EM) field. Here and in the following, we use atomic units
($\hbar=4\pi\epsilon_0=m_e=1$) unless stated otherwise.
\begin{multline}\label{eq:HQED}
\hat{H} = \sum_{i} \frac{(\vecop{p}_i-Q_i\vecop{A}(\br_i))^2}{2m_i} + \sum_{i>j} \frac{Q_i Q_j}{|\br_{i} - \br_j|} \\
+ \frac{1}{8\pi} \int \left(\vecop{E}^\perp(\br)^2 + c^2 \vecop{B}(\br)^2\right) \mathrm{d}^3\br,
\end{multline}
where $\vecop{E}^\perp(\br)=-\frac1c \frac{\partial\vecop{A}}{\partial t}$ is
the transverse part of the electric field (with the longitudinal part
responsible for the instantaneous Coulomb interaction $Q_i Q_j/r_{ij}$), and we
use the Coulomb gauge $\nabla\cdot\vecop{A} = 0$. We note explicitly that here,
the EM operators represent free-space modes (i.e., without boundary conditions
imposing a cavity structure), while the collection of charged particles
(specifically, electrons and nuclei) represent both the material part of the
cavity (e.g., mirrors) and the emitters (such as molecules). In particular, the
cavity material together with the EM field modes will have approximately bosonic
eigenmodes that can be identified as the ``cavity modes'' and, in general, will
be given by superpositions of material and EM field
excitations~\cite{Scheel2008}, as explicitly shown for plasmonic systems
in~\cite{Alpeggiani2014}. For simplicity and generality, in the following we
assume that the cavity-molecule system we are treating is well-described within
the quasistatic approximation, which applies when all distances in the problem
are significantly smaller than the relevant wavelengths. In particular, this is
a good approximation for small plasmon- and phonon-polariton nanoantennas and
nanoresonators, which are the only currently available systems that achieve
strong enough field concentration to obtain strong single-emitter couplings with
``real'' atoms or molecules~\cite{Zengin2015, Chikkaraddy2016, Chikkaraddy2018,
Autore2017, Gubbin2017Theoretical} (as opposed to ``artificial atoms'' such as
superconducting qubits~\cite{Niemczyk2010,Forn-Diaz2016,Yoshihara2017}). In the
quasistatic limit, the transversal fields are negligible\footnote{More
precisely, the weak coupling to transversal fields still induces free-space QED
effects such as the Lamb shift and radiative decay, but they are not
significantly modified by the presence of the cavity.}, so that
$\vecop{A}=\vecop{B}=\vecop{E}^\perp\approx0$, and the Hamiltonian simply
becomes
\begin{equation}\label{eq:HQEDqs}
	\hat{H} = \sum_{i} \frac{\vecop{p}_i^2}{2m_i} + \sum_{i>j} \frac{Q_i Q_j}{|\br_{i} - \br_j|},
\end{equation}
with the sums over $i$ and $j$ still including all particles in the (nano)cavity
as well as the molecules. We next separate the particles into several groups:
one containing the cavity material, and one for each molecule. We assume that
the cavity material is ``macroscopic'' enough that it responds linearly to
external fields~\cite{Hopfield1958, Huttner1992, Dung1998, Scheel2008,
VanVlack2012, Alpeggiani2014}, and can thus be well-described by a collection of
bosonic modes with frequencies $\omega_k$ and annihilation operators $a_k$
(e.g., corresponding to the ``instantanteous'' plasmon modes
in~\cite{Alpeggiani2014}). For simplicity, we first consider a single molecule
including $n_\mathrm{e}$ electrons and $n_\mathrm{n}$ nuclei. The Hamiltonian
then becomes
\begin{multline}\label{eq:H}
\hat{H} = \sum_{i=1}^{n_\mathrm{n}} \frac{\vecop{P}_i^2}{2M_i} +
	\hat{H}_\mathrm{e}(\vecop{x},\vecop{R}) + \sum_k \omega_k \hat{a}_k^\dagger \hat{a}_k \\
	+ \sum_k (\hat{a}_k+\hat{a}_k^\dagger) \sum_j Q_j \phi_k(\vecop{r}_j).
\end{multline}
The bare molecular Hamiltonian corresponds to the first two terms: the kinetic
energy of $n_\mathrm{n}$ nuclei and the electronic Hamiltonian. The latter
includes the kinetic energy of the $n_\mathrm{e}$ electrons and the
nucleus-nucleus, electron-electron, and nucleus-electron interaction potentials.
This operator depends on all the electronic and nuclear positions,
$\vecop{x}=(\vecop{x}_1,\vecop{x}_2,\dots,\vecop{x}_{n_\mathrm{e}})$ and
$\vecop{R}=(\vecop{R}_1,\vecop{R}_2,\dots,\vecop{R}_{n_\mathrm{n}})$,
respectively. The following two terms correspond to the bosonic cavity modes and
the interaction of the molecular charges (with $j$ running over both electrons
and nuclei) with the electrostatic potential $\phi_k(\br)$, i.e., the Coulomb
potential corresponding to the charge distribution of each cavity mode. By
performing a multipole expansion of the molecular charges, and assuming that the
molecule is uncharged and sufficiently localized, this term can be
well-approximated by $\vecopsym{\mu}\cdot\vecop{E}(\br_m)$, i.e., the
interaction of the molecular dipole with the cavity electric field (the gradient
of the potential) at the position $\br_m$ of the molecule, which we write as
\begin{equation}\label{eq:Hint}
	(\hat{a}_k+\hat{a}_k^\dagger) \sum_j Q_j \phi_k(\vecop{r}_j)
	\approx \omega_k \hat{q}_k \blambda_k \cdot \vecopsym{\mu}(\vecop{x},\vecop{R}),
\end{equation}
where $\hat{q}_k = \frac{1}{\sqrt{2\omega_k}} (\hat{a}_k + \hat{a}_k^\dagger)$
is the position operator of the harmonic oscillator, and the electric field
strength is determined by $\blambda_k = \lambda_k \boldsymbol{\epsilon}_k$, with
polarization vector $\boldsymbol{\epsilon}$. The coupling constant can be
related to both the single-photon electric field strength and the
(position-dependent) effective mode volume of the quantized mode, with
$\lambda_k = \sqrt{\frac{2}{\omega_k}} E_{\mathrm{1ph},k}(\br_m) =
\sqrt{4\pi/V_{\mathrm{eff},k}}$. Here, the effective EM mode volume is defined
as $V_{\mathrm{eff},k} = \frac{\int\varepsilon(\br)^2|\textbf{E}(\br)|^2
\mathrm{d}^3\br} {\varepsilon(\br_m)^2|\textbf{E}(\br_m)|^2}$, although the
normalization integral formally diverges for lossy modes and has to be properly
generalized~\cite{Koenderink2010,Kristensen2012,Sauvan2013,Alpeggiani2014}. 

The proper description of the light-matter interaction Hamiltonian under
(ultra)strong-coupling conditions is a very active topic of discussion in the
literature~\cite{Vukics2014, DeBernardis2018Breakdown, DeBernardis2018Cavity,
Rokaj2018, Stokes2019, Andrews2018, Konya2018, Rousseau2018, SanchezMunoz2018}.
In particular, much of this discussion centers on the importance of the
so-called dipole self-energy term $\frac12(\blambda \cdot
\vecopsym{\mu}(\vecop{x},\vecop{R}))^2$ that arises in the
Power--Zienau--Woolley transformation, where the interaction with the
transversal field $\vecop{A}$ is transformed to an electric-field--dipole
interaction with the same form as \autoref{eq:Hint} plus the above-mentioned
dipole self-energy term. As we have discussed, and as is well-known in the
literature on macroscopic QED~\cite{Buhmann2007Thesis}, this term does not
appear for interaction with purely longitudinal modes that are well-described
within the quasistatic approximation, i.e., in situations where retardation and
propagation effects of the EM fields can be neglected. Given the fact that
reaching strong or ultrastrong coupling with one (or a few) atoms or molecules
requires strongly sub-wavelength mode volumes, $V_{\mathrm{eff},k} \ll (2\pi
c/\omega_k)^3$, it follows that the quasistatic approximation should be
applicable for most realistic cavities with few-emitter strong coupling. On the
other hand, this extreme field localization also can require going beyond the
point-dipole interaction either by directly using the interaction with the full
space-dependent potential $\phi_k(\mathbf{r})$~\cite{Neuman2018Coupling} or by
including higher multipoles in \autoref{eq:Hint}~\cite{Cuartero-Gonzalez2018}.
Doing so also resolves the formal lack of a ground state when the computational
box is made too large and no dipole self-energy term is
present~\cite{Rokaj2018,Schafer2018}.

However, it should be noted that if the sum over cavity modes is truncated and
the effect of all but one (or a few) modes is approximately represented by
renormalizing the emitter potential (and emitter-emitter interactions in the
multiple-emitter case), it is necessary to add back an effective (collective)
dipole self-interaction to avoid double-counting of modes, as explained
in~\cite{DeBernardis2018Cavity}.

We note that while we have explicitly treated a (nano)cavity within the
quasistatic approximation, in which the cavity fields can be understood as due
to the instantaneous Coulomb interaction between charged particles, it still
makes sense to speak of the cavity modes as electromagnetic or photonic modes
with an associated electric field. The modes, which physically correspond to,
e.g., plasmonic or phonon-polaritonic resonances, can be seen as strongly
confined photons. These modes are most easily obtained by solving Maxwell's
equations for a given geometry, either numerically or with approaches such as
transformation optics~\cite{Li2016Transformation}. Only in the limit of
extremely small nanocavities does it become possible, and sometimes necessary,
to treat them explicitly as a collection of nuclei and electrons using ab initio
techniques~\cite{Savage2012,Zhang2014,Varas2016}.

In the following, we will first treat a cavity in which only a single mode has
significant coupling to the molecule (in \appref{app:sphere_quantization}, we
discuss some systems in which this is a valid approximation). Since the
interaction depends on the inner product between the electric field and the
total dipole moment $\vecopsym{\mu} = \sum_i^{n_\mathrm{n}} Z_i \vecop{R}_i -
\sum_i^{n_\mathrm{e}}
\vecop{x}_i$, only the projection $\hat{\mu}_\epsilon = \vecopsym{\epsilon}
\cdot \vecopsym{\mu}$ is relevant, and we only have to deal with scalar
quantities. For the sake of simplicity, we rewrite $\hat{\mu}_\epsilon
\rightarrow \hat{\mu}$. We also assume perfect alignment between the molecule
and the field unless indicated otherwise.


\subsection{Molecular model}

\begin{figure}
\includegraphics[width=\linewidth]{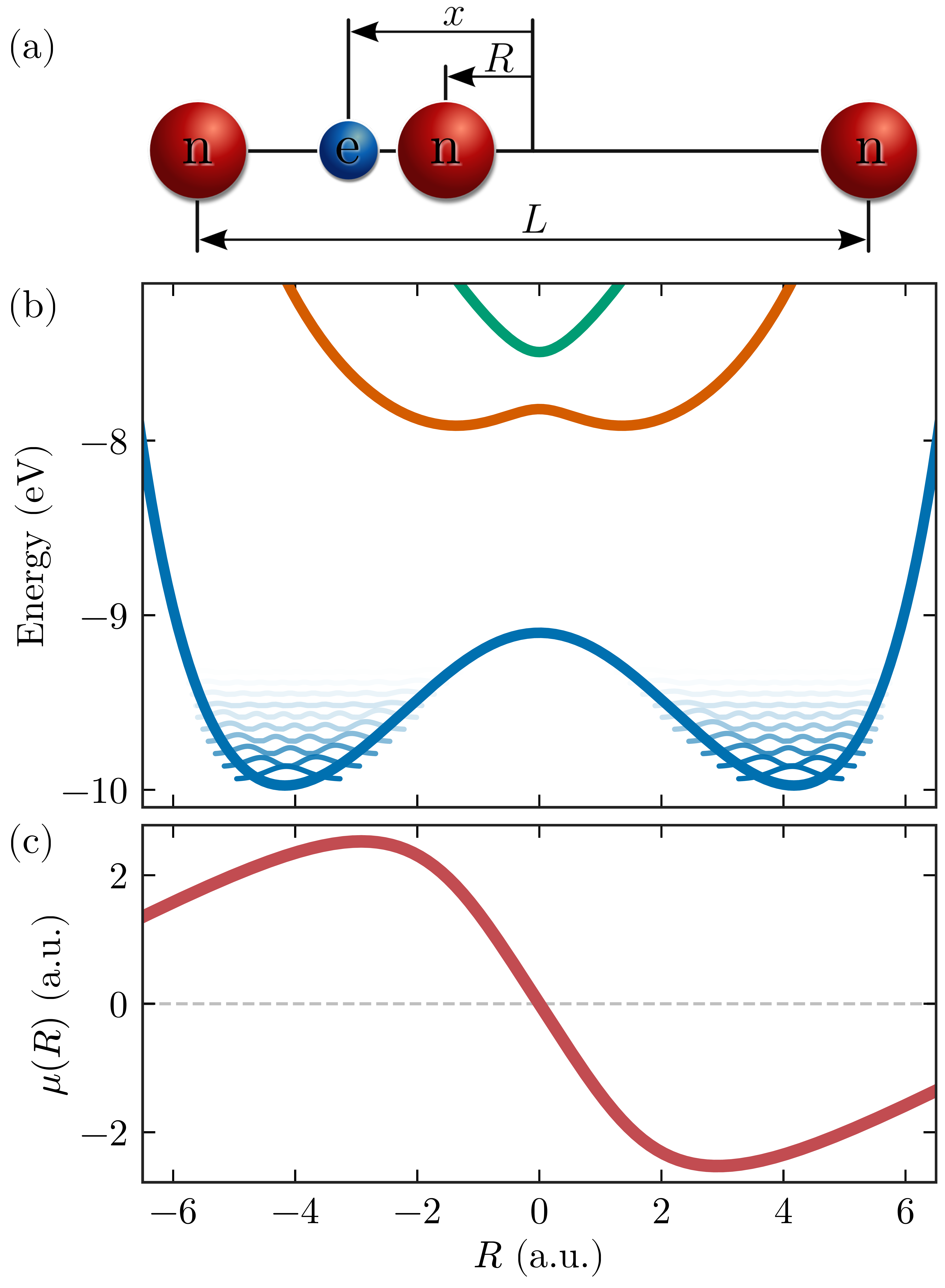}
\caption{(a) Schematic representation of the Shin--Metiu model close to one of the
equilibrium configurations. The two ions on both sides are fixed at a distance
$L$, while the electron and the remaining ion can move freely in between. (b)
Potential energy surfaces of the model with the vibrational levels and
associated probability densities of the ground state (blue) represented. (c)
Ground state dipole moment.}
\label{fig:shinmetiu}
\end{figure}

In order to study changes in ground-state chemical reactivity induced by
(vibrational) strong coupling, we first treat a simple molecular model system
that is numerically fully solvable and has been extensively studied in model
calculations of chemical reaction rates, the Shin--Metiu model~\cite{Shin1995}.
It treats three nuclei and one electron moving in one dimension, as presented in
\autoref{fig:shinmetiu}(a). Two of the nuclei are separated by a distance $L$
and fixed in place, while the remaining nucleus and the electron are free to
move. The repulsive interaction of the mobile nucleus with the fixed ones is
given by a normal Coulomb potential, while the attractive electron-nuclei
interaction is given by softened Coulomb potentials
$V_\mathrm{en}(r_i)=Z\mathrm{erf}(r_i/R_c)/r_i$, where $r_i$ is the distance
between the electron and nucleus $i$ and $R_c$ is the softening parameter. The
system has two stable nuclear configurations (minima of the ground-state
Born--Oppenheimer surface) that represent two different isomers of a charge or
proton transfer reaction. Given that the electronic excitations energies and
thus the nonadiabatic couplings between different potential energy surfaces can
be varied easily by changing the parameters of the Shin--Metiu model, it has been
extensively studied in the context of correlated electron-nuclear
dynamics~\cite{Shin1996,Abedi2013}, as well as in the context of polariton
formation under strong coupling~\cite{Flick2017Atoms, Flick2017Cavity}. The
parameters chosen throughout the present work are $Z=1$, $L=10$~\AA~$\approx
18.9$~a.u., $M=1836$~a.u., and $R_c=1.5$~\AA~$\approx 2.83$~a.u.\ (for all three
nuclei), resulting in the Born--Oppenheimer potential energy surfaces shown in
\autoref{fig:shinmetiu}(b), with negligible nonadiabatic coupling between
electronic surfaces. The figure also shows the first few vibrational eigenstates
close to each minimum (tunneling through the central energy barrier is
negligible for these states, so that they can be chosen to be localized on the
left or right, respectively). In \autoref{fig:shinmetiu}(c) we show the
ground-state permanent dipole moment $\mu_g(R) = \langle g|\mu(R)|g\rangle$.
Below we demonstrate that, to a good approximation, the ground-state potential
energy surface and dipole moment are sufficient to describe the change in the
molecular ground-state structure and chemical reactivity due to the cavity.
Additionally, we note here that the light-matter coupling strength for formation
of vibro-polaritons, i.e., hybridization of the photon mode with the vibrational
transitions of the molecule, is determined by the transition dipole moment and
frequency of the quantized vibrational levels of the molecule. Within a
lowest-order expansion around the equilibrium position, $V_g(R)\approx V_g(R_0)
+ \frac{1}{2} M \omega_\nu^2 (R-R_0)^2$, $\mu_g(R)\approx\mu_g(R_0) +
\mu'_g(R_0)(R-R_0)$, these are given by $\omega_\nu = 72.6$~meV, and $\mu_v
\approx \frac{1} {\sqrt{2M\omega_\nu}} \mu'_g(R_0)$, giving a Rabi frequency
$\Omega_R = \frac{\lambda}{\sqrt{M}} \mu'_g(R_0)$ on resonance ($\omega_c =
\omega_\nu$)~\cite{Shalabney2015}.


\section{Quantum reaction rates}\label{sec:rates}

In this section, we analyze the cavity-induced change in the rate of the
ground-state proton-transfer reaction from the left minimum at $R\approx
-4$~a.u.\ to the right one. In the present section, we take advantage of the
simplicity of the Shin--Metiu model to exactly compute the quantum reaction rate
without any approximations, which automatically takes into account all quantum
effects such as tunneling or zero-point energy. We follow the approach of
Miller~\cite{Miller1983}, based on the correlation function formalism introduced
in~\cite{Yamamoto1960,Miller1974}. This states that the rate for a molecular
reaction is given by
\begin{equation}\label{eq:rate}
k(T) = \frac{1}{Q_\mathrm{r}(T)} \int_0^{t_f\rightarrow \infty} C_{ff}(t) \mathrm{d}t,
\end{equation}
where $Q_\mathrm{r}(T) = \mathrm{tr} [\exp{(-\beta \hat{H})}]$, with $\beta^{-1}
= k_\mathrm{B}T$, is the partition function of the reactants at temperature $T$
and $C_{ff}(t)$ is the flux-flux autocorrelation function, defined as
\begin{equation}\label{eq:Cff}
C_{ff}(t) = \mathrm{tr}[\bar{F}\hat{U}^\dagger(t_c) \bar{F} \hat{U}(t_c)].
\end{equation}
This correlation function is computed as the trace of a product of operators,
where $U(t_c) = \exp(-i \hat{H} t_c)$, with $t_c=t-i\beta/2$, is the complex
time evolution operator and $\bar{F}$ represents the symmetrized flux operator
\begin{equation}\label{eq:flux}
\bar{F} = \frac{1}{2M}\left( \hat{P} \delta(s) + \delta(s) \hat{P} \right).
\end{equation}
Here, $\hat{P}$ is the nuclear momentum operator and the surface dividing the
reactant and product states is defined by the zeros of the function $s=s(R)$. In
our case, the line that defines products and reactants is $R=0$, i.e., $s(R)=R$.
The flux-flux autocorrelation function describes the temporal flux of
positive-momenta probability through the dividing surface of a thermally
averaged initial state (which is accounted for by the thermal part of the
$\hat{U}(t_c)$ operator). Negative values of $C_{ff}(t)$ indicate recrossing of
the dividing surface in the opposite direction, thus contributing to a rate
decrease.

In order to obtain the rates of the coupled electronic-nuclear-photonic system,
we discretize all three degrees of freedom, using a finite-element discrete
variable representation~\cite{SchFeiNag2011} for $x$ and $R$, as well as the
Fock basis for the cavity photon mode. This allows to diagonalize the full
Hamiltonian, \autoref{eq:H}, and thus to trivially calculate \autoref{eq:Cff}
for arbitrary time $t$. For numerical efficiency, we perform the diagonalization
in steps, first diagonalizing the bare molecular Hamiltonian, performing a
cut-off in energy, and then diagonalizing the coupled system in this basis. We
have carefully checked convergence with respect to all involved grid and basis
set parameters and cutoffs. As is well known~\cite{Shin1995}, due to the absence
of dissipation in the model, for large times the correlation function becomes
negative and oscillates around zero, corresponding to the wave packet that has
crossed the barrier returning back through the dividing surface after reflection
at the other side of the potential (at $R\approx 6$~a.u.). However, in a real
system the reaction coordinate is coupled to other vibrational and solvent
degrees of freedom that will dissipate the energy and prevent recrossing. To
represent this, we choose a final time $t_f$ around which the correlation
function stays equal to zero for a while and only integrate up that time in
\autoref{eq:rate}. The time chosen, $t_f = 35$~fs, corresponds to typical
dissipation times in condensed phase reactions, and is similar to values chosen
in the cavity-free case~\cite{Shin1995}.

\begin{figure}
	\includegraphics[width=\linewidth]{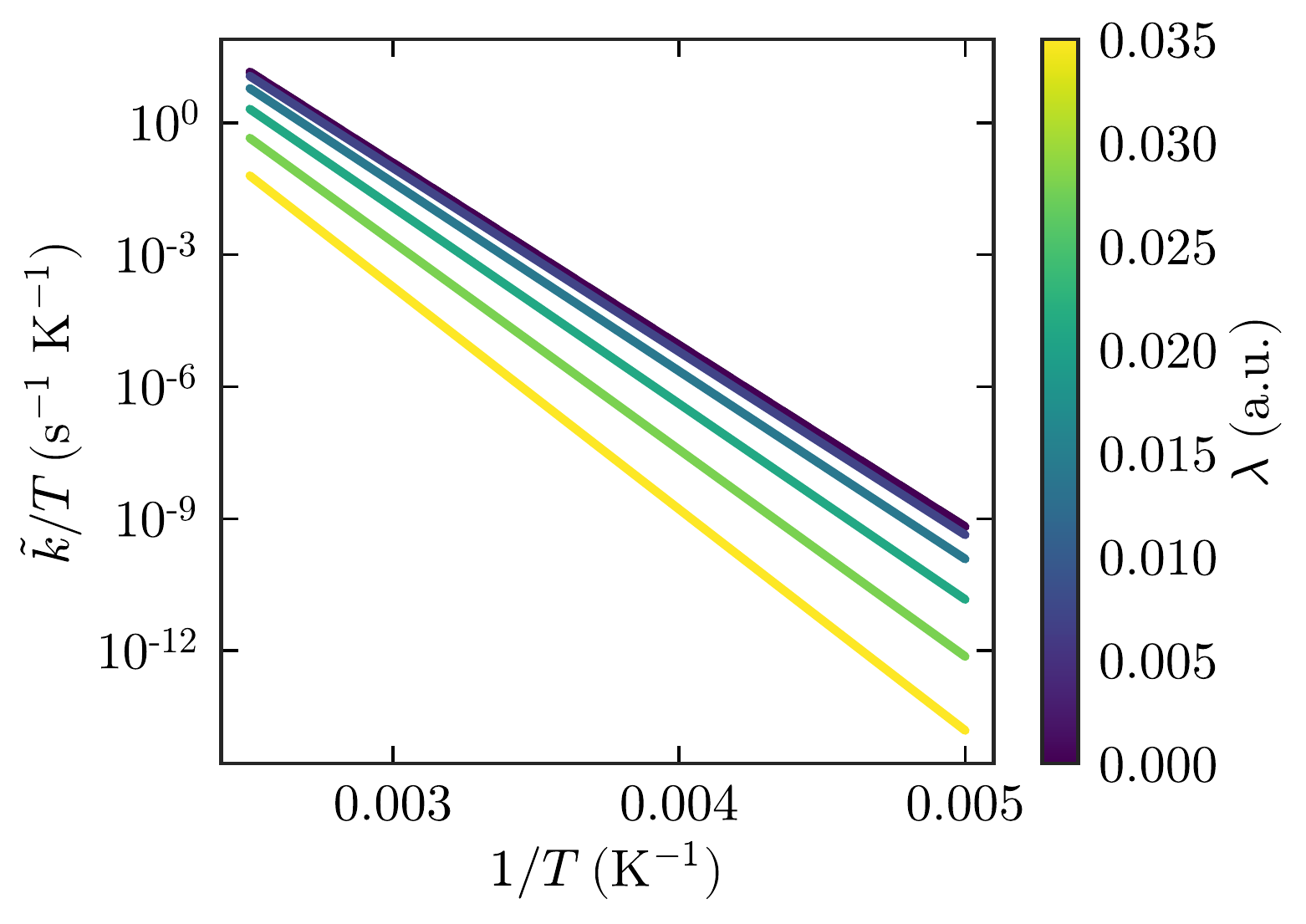}
	\caption{Arrhenius plot for the rate dependence with temperature in the
	hybrid system for several light-matter coupling values. See main text for
	details.}
	\label{fig:eyring}
\end{figure}

We now study the cavity-modified chemical reaction rates of the hybrid system
for different coupling strengths $\lambda$. We note that a coupling strength of
$\lambda = 0.035$~a.u.\ corresponds to a Rabi splitting of $\Omega_\mathrm{R}
\approx 0.10 \omega_\nu$ for the first vibrational transition. For the sake of
comparison, we mention that single-molecule electronic strong coupling has been
achieved with mode volumes of $\sim 40$~nm$^3$~\cite{Chikkaraddy2016},
corresponding to $\lambda \approx 0.007$~a.u., and there are indications that
effective sub-nm$^3$ mode volumes could be reached due to single-atom hot
spots~\cite{Benz2016,Urbieta2018}, which would allow the coupling strength to
reach values up to $\lambda \approx 0.05$~a.u.. \autoref{fig:eyring} shows the
rates in an Arrhenius plot, i.e., the logarithm of the rate divided by the
temperature as a function of the inverse temperature. The straight lines in
\autoref{fig:eyring} confirm that the hybrid light-matter system follows the
behavior described by the Eyring equation~\cite{Eyring1935}, which connects the
rate of a chemical reaction with the energy barrier $E_\mathrm{b}$ that
separates reactants from products:
\begin{equation}\label{eq:TST}
k= \kappa 2\pi k_\mathrm{B} T e^{-\frac{E_\mathrm{b}}{k_\mathrm{B}T}}.
\end{equation}
Here, $\kappa$ is a transmission coefficient, typically considered equal to one
if nonadiabatic effects can be neglected close to the transition state. This
equation follows from classical transition state theory~\cite{Eyring1935,
Laidler1987} and is often used in the context of chemical kinetics.

\begin{figure*}
	\includegraphics[width=\linewidth]{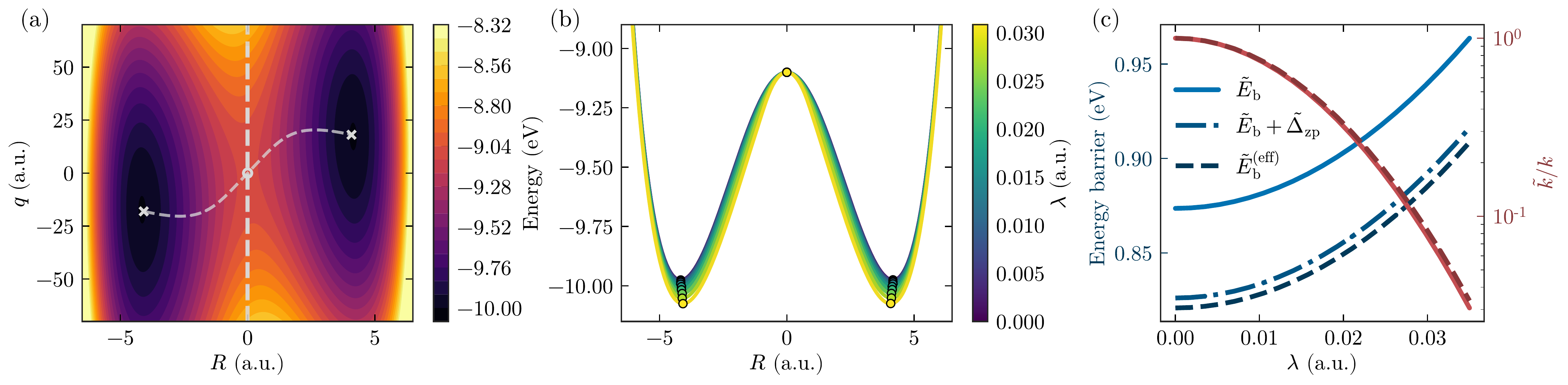}
	\caption{(a) Two-dimensional ground-state PES in the cavity
	Born--Oppenheimer approximation for the Shin--Metiu model for $\lambda =
	0.02$~a.u.\ and $\omega_c=72.6$~meV. At $R=0$ we show the dividing surface
	used to compute the reaction flux from reactant to product states. The gray
	dashed line curve corresponds to the energy path along $q_\mathrm{m}(R)$,
	i.e., the minimum in $q$. (b) Value of the energy path $V_0(\bR,
	q_\mathrm{m})$ for different values of the Rabi frequency, which is related
	to the coupling strength through $\Omega_\mathrm{R} = \lambda \mu'_0(R_0)$,
	where the dipole derivative is evaluated at the minimum, as discussed in
	\appref{app:normal_modes}. (c) Energy barrier and rates ratio vs
	coupling strength for the case of a CBOA calculation (full lines) and for
	the effective energy barrier fitted from exact quantum rate calculations
	(dashed lines).}
	\label{fig:CBOA}
\end{figure*}

We thus observe that even under vibrational strong coupling and the accompanying
formation of vibro-polaritons, i.e., hybrid light-matter excitations, the
reaction rate can still be described by an effective potential energy barrier.
However, the effective height of the energy barrier is modified through the CQED
effect of strong coupling, leading (for the studied model) to significantly
reduced reaction rates. Although we here treat a single-mode and single-molecule
system, these general observations agree with experimental
studies~\cite{Thomas2016, Thomas2019, Hiura2018}. However, in order to gain
further insight into this effect and enable calculations beyond simple model
systems, it would be desirable to have a theory that is not based on full
quantum rate calculations (which require the calculation of nuclear dynamics in
$3N-6$ dimensions). In the next section, we show that this can be achieved by
applying (classical) transition state theory to the combined photonic-nuclear
potential energy surfaces provided by the Cavity Born--Oppenheimer
approximation~\cite{Flick2017Cavity}.


\section{Cavity Born--Oppenheimer approximation}\label{sec:CBOA}

The starting point of the cavity Born--Oppenheimer approximation
(CBOA)~\cite{Flick2017Atoms,Flick2017Cavity} is to write the cavity mode energy
as an explicit harmonic oscillator,
\begin{equation}
	\omega_c \left(\hat{a}^\dagger\hat{a} + \frac12\right) = \frac{\hat{p}^2}{2} + \omega_c^2 \frac{\hat{q}^2}{2},
\end{equation}
with $\hat{p} = i\sqrt{\frac{\omega_c}{2}}(\hat{a}^\dagger - \hat{a})$ and
$\hat{q} = \frac{1}{\sqrt{2\omega_c}} (\hat{a}^\dagger + \hat{a})$ as discussed
in \autoref{sec:theory}. The cavity photon degree of freedom is then treated as
nuclear-like and its ``kinetic energy'' $\hat{p}^2/2$ grouped with the nuclear
kinetic energy operators $\sum_i \vecop{P}_i^2 / (2M_i)$ before performing the
standard Born--Oppenheimer approximation. This leads to a set of electronic CBO
potential energy surfaces (PES) $\tilde V_i(\bR,q)$ parametric in both
nuclear $\bR$ and photonic coordinates $q$, obtained by diagonalizing the
new electronic Hamiltonian, $\hat{H}_\mathrm{e} (\vecop{x};\bR,q) =
\hat{H} - \frac{\hat{p}^2}{2} - \sum_{i=1}^{n_\mathrm{n}}
\frac{\vecop{P}_i^2}{2M_i}$:
\begin{equation}\label{eq:schrodinger}
\hat{H}_\mathrm{e} (\vecop{x};\bR,q) \phi_i(\textbf{x};\bR,q) = \tilde{V}_i(\bR,q)\phi_i(\textbf{x};\bR,q).
\end{equation}
Conceptually, the inclusion of the cavity mode thus simply corresponds to a
single additional nuclear-like degree of freedom.

The CBOA now consists in neglecting nonadiabatic couplings between different PES
(i.e., neglecting the action of nuclear and photonic kinetic operators on the
electronic states) and assuming photonic and nuclear ``motion'' to proceed on
each PES independently. Due to the formal equivalence between nuclear and
photonic degrees of freedom within this picture, all the standard results of BO
theory apply. In particular, the CBOA is a good approximation when the
separation between the PES is larger than typical kinetic energies of the nuclei
and the photonic mode. The case of vibrational strong coupling, where the photon
energy is comparable to vibrational excitation energies, exactly fulfills this
condition. The accompanying Rabi splitting can then be understood as simply
normal mode hybridization on the nuclear-photonic potential energy surface, as
already noted in the original article demonstrating vibrational strong
coupling~\cite{Shalabney2015}, and discussed in more detail in
\appref{app:normal_modes}.

In the context of cavity-modified chemical reactivity in the ground state, the
formal equivalence between photonic and nuclear motion in the CBOA in particular
allows to apply standard tools such as transition state theory to obtain an
estimation for reaction rates. TST implies that it should only be necessary to
calculate the effective energy barrier for the reaction within the ground-state
CBO surface.

We test this for the model studied in \autoref{sec:rates}, i.e., the Shin--Metiu
model coupled to a cavity mode on resonance with the first vibrational
transition. The two-dimensional PES $\tilde V_0(R,q)$ is shown in
\autoref{fig:CBOA}(a) for a coupling strength of $\lambda = 0.02$~a.u., which
corresponds to a vibrational Rabi splitting of $\Omega_\mathrm{R} \approx 0.05
\omega_\nu$. The second panel, \autoref{fig:CBOA}(b), shows the minimum along
$q$ of this surface as a function of $R$, i.e., along the path indicated by the
curved dashed line in \autoref{fig:CBOA}(a), for a set of coupling strengths
$\lambda$ that induce a Rabi splitting of up to $\Omega_R = 0.1 \omega_\nu$.
This path closely corresponds to the minimum energy path of the proton transfer
reaction within the CBOA\@. As the coupling is increased, the minima become
deeper, while the transition state (TS) at $R=0$ stays unaffected. This leads to
an effective increase of the reaction barrier $\tilde{E}_\mathrm{b} =
\tilde{V}_0(R_\mathrm{TS},q_\mathrm{TS}) -
\tilde{V}_0(R_\mathrm{min},q_\mathrm{min})$, as shown in \autoref{fig:CBOA}(c).
This panel also shows the corresponding change in the rate predicted by
\autoref{eq:TST}. The full lines correspond to the energy barrier calculated
within the CBOA (blue) and the corresponding rate (red) according to TST, while
the dashed lines show the effective energy barrier
$E_\mathrm{b}^{(\mathrm{eff})}$ extracted from the fit to the Arrhenius plot
\autoref{fig:eyring} and the corresponding change in the rate obtained from the
full quantum rate calculation in \autoref{sec:rates}. As can be seen, the
effective and CBOA energy barriers agree very well, with just an approximately
constant overestimation of the barrier in CBOA due to quantum effects such as
zero-point energy and tunneling. This leads to excellent agreement for the
change of the reaction rate obtained from the full quantum calculation and the
CBOA-TST prediction. As expected from our previous discussion, the reaction rate
of the hybrid cavity-molecule system decreases dramatically as the coupling
increases due to the increase of the energy barrier height. Finally, we also
calculate the CBOA energy barrier corrected by $\tilde{\Delta}_\mathrm{zp}$, the
difference between the zero-point vibrational frequencies at the minimum and
transition states as obtained from the Hessian of the PES (disregarding the
direction of negative curvature at the TS). This is shown as a dash-dotted line
in \autoref{fig:CBOA}(c), and considerably improves the absolute agreement with
the effective barrier extracted from the full quantum rate calculations.

While we have up to now worked within a single-mode model, the CBO approximation
actually makes it straightforward to treat multiple photonic modes. The ground
state PES then parametrically depends on multiple parameters $q_k$, one for each
mode, just as a realistic molecule depends on multiple nuclear positions
$\bR_i$. Similarly, the adiabatic surfaces are not harder to calculate
than for the single-mode case, and minimization strategies can rely on the same
approaches used in ``traditional'' quantum chemistry. We note that for a general
cavity, the mode parameters can be obtained either by explicitly quantizing the
modes (which is in general a difficult proposition) or, alternatively, by
rewriting the spectral density of the light-matter coupling (proportional to the
EM Green's function) as a sum of Lorentzians~\cite{Gonzalez-Tudela2014,
Delga2014, Delga2014a, Li2016Transformation}.


\section{Perturbation Theory}\label{sec:pert_th}

As we have seen, the cavity Born--Oppenheimer approximation provides a
convenient picture to evaluate cavity-induced changes in chemical reactivity
based on energy barriers in electronic PES that are parametric in nuclear and
photonic coordinates. In particular, the interaction term $\omega_c q
\blambda\cdot \vecopsym{\mu}$, with $q$ a parameter, is equivalent
to that obtained from applying a constant external electric field. The CBO PES
for arbitrary molecules can thus be calculated with standard quantum chemistry
codes. However, obtaining the barrier in general still requires minimization of
the molecular PES along the additional photon coordinate $q$ (or coordinates
$q_k$, if multiple modes are treated). If the coupling is not too large and the
relevant values of $q$ are small enough, the CBO ground-state PES can instead be
obtained within perturbation theory, which up to second order in $\lambda$ is
given by
\begin{equation}\label{eq:VgsSC}
\tilde{V}_0(\bR,q) \approx V_0(\bR) + \frac{\omega_c^2}{2} q^2 +
	\lambda \omega_c q \mu_0(\bR) - \frac{\lambda^2}{2}\omega_c ^2 q^2 \alpha_0(\bR),
\end{equation}
where $V_0(\bR)$ and $\mu_0(\bR)$ are the bare-molecule
ground-state PES and dipole moment, respectively, while $\alpha_0(\bR)$
is the ground-state static polarizability~\cite{Bonin1997},
\begin{equation}
\alpha_0(\bR;\omega=0)=2\sum_{m\neq0}\frac{|\mu_{m,0}(\bR)|^2}{V_m(\bR)-V_0(\bR)},
\end{equation}
and encodes the effect of excited electronic levels, with
$\mu_{m,0}(\bR)$ the transition dipole moment between bare-molecule
electronic levels $m$ and $0$. Obtaining the full ground-state CBO surface
within this approximation then just requires the calculation of the
bare-molecule ground-state properties $V_0(\bR)$, $\mu_0(\bR)$, and
$\alpha_0(\bR)$.

\begin{figure}
	\includegraphics[width=0.9\linewidth]{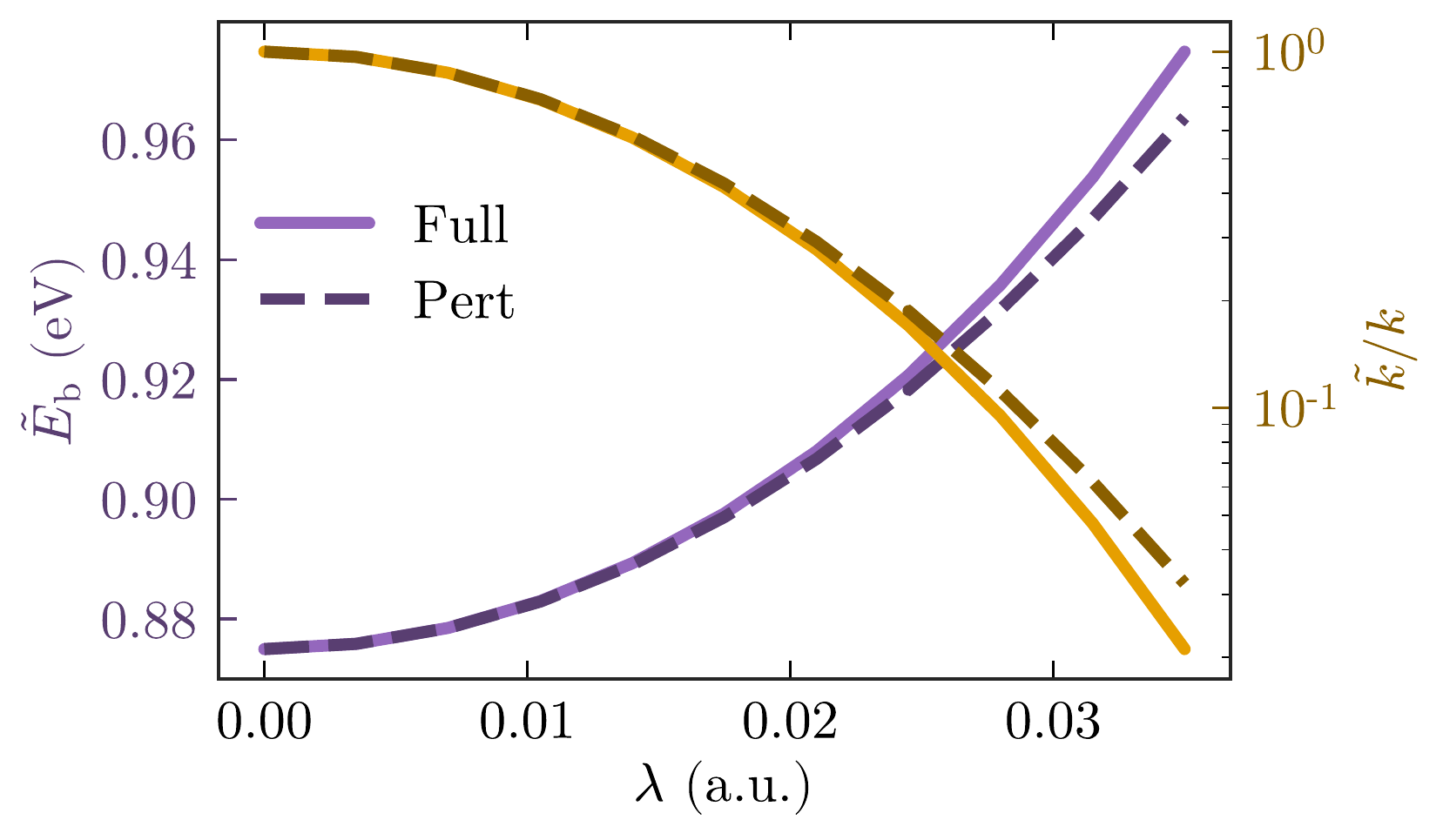}
	\caption{Cavity Born--Oppenheimer energy barrier (purple) and relative
	change of reaction rates (yellow) for the Shin--Metiu model inside a cavity,
	calculated to all orders in the light-matter coupling strength $\lambda$
	(solid lines), and up to second order in perturbation theory (dashed
	lines).}
	\label{fig:pert_validity}
\end{figure}

In addition to providing an explicit expression for the CBO ground-state PES in
terms of bare-molecule ground-state properties, the simple analytical dependence
on $q$ in \autoref{eq:VgsSC} allows to go one step further and obtain explicit
expressions for the local minima and saddle points (i.e., transition states). In
these configurations, the conditions $\partial_q \tilde{V}_0(\bR,q)=
\partial_\bR \tilde{V}_0(\bR,q)=0$ are satisfied. This yields a
set of coupled equations that can be solved in order to find the configuration
of the new critical points along the reaction path. The first equation gives the
explicit condition
\begin{equation}\label{eq:qmin}
	q_\mathrm{m}(\bR) = -\frac{\lambda}{\omega_c}
		\frac{\mu_0(\bR)}{1-\lambda^2\alpha_0(\bR)},
\end{equation}
which can be used to obtain the potential profile along the minimum in $q$,
\begin{equation}\label{eq:V_at_qm}
	\tilde{V}_0(\bR,q_m) = V_0(\bR) - \frac{\lambda^2}{2}\mu_0^2(\bR) +
	\mathcal{O}(\lambda^4),
\end{equation}
where we have dropped terms of order $\lambda^4$ since the perturbation-theory
PES \autoref{eq:VgsSC} is only accurate to second order. This shows that the
energy barrier on the CBO surface (within second-order perturbation theory) can
be calculated directly from the bare-molecule potential and permanent dipole
moment. In \autoref{fig:pert_validity}, we analyze the validity of
\autoref{eq:V_at_qm} for computing the barrier height within the Shin--Metiu
model. It can be observed that perturbation theory works quite well for the
whole range of couplings, with a relative error in the cavity-induced change of
the energy barrier of about $10\%$ for the largest considered couplings. Due to
the exponential dependence of the rates on barrier height, this corresponds to
an appreciable error in the rate constant, but still provides a reasonable
estimate. Note that in the case of the Shin--Metiu model, the error of the
energy barrier stems entirely from the change at the minimum configuration, as
the transition state has zero dipole moment due to symmetry and is not affected
by the cavity.

It is interesting to point out that \autoref{eq:V_at_qm} closely resembles the
expression obtained in electric field catalysis where an external voltage is
applied~\cite{Fried2017}, or to electrostatic shifts provided by some
catalysts~\cite{Welborn2018}. This strategy exploits the Stark effect, i.e., the
energy shift observed in the presence of a static electric field, to induce
changes in the energies of the transition state relative to the minimum
configuration. As noted before, the CBOA corresponds to treating the influence
of the cavity through an adiabatic parameter $q$ determining the electric
field strength. However, instead of being externally imposed, in our case the
effective field, determined by \autoref{eq:qmin}, is the one induced in the
cavity by the permanent dipole moment of the molecule itself. This also lends
itself to an electrostatic interpretation of the effect.

In addition to the minimum energy barrier of the CBO PES itself, the effective
energy barrier is also affected by the zero-point energy due to the quantization
of nuclear and photonic motion (see \autoref{fig:CBOA}). We can obtain its
cavity-induced shift within perturbation theory by using \autoref{eq:qmin} to
rewrite \autoref{eq:VgsSC} as
\begin{equation}\label{eq:V_around_qm}
	\tilde{V}_0(\bR,q) = \tilde{V}_0(\bR,q_m) +
	\frac{\omega_{\mathrm{eff}}^2(\bR)}{2} (q - q_m(\bR))^2,
\end{equation}
where $\omega_{\mathrm{eff}}(\bR) = \omega_c - \frac{\lambda^2}{2}\omega_c
\alpha_0(\bR) + \mathcal{O}(\lambda^4)$, such that the photonic zero-point
energy $\omega_{\mathrm{eff}}(\bR)/2$ is decreased due to the polarizability of
the molecule. We note that this only accounts for the quantization of the
photonic motion along $q$. As we show in \appref{app:normal_modes},
close to a local minimum at $R_0$, there is an additional correction due to the
vibrational contribution to the molecular polarizability, which to second order
is given by $-\frac{\omega_c \Omega_R^2}{4\omega_v (\omega_c +\omega_v)}$, where
$\Omega_R = \frac{\lambda}{\sqrt{M}} \mu'_g(R_0)$ is the on-resonance
vibrational Rabi splitting as discussed in \autoref{sec:theory}. As can be
appreciated from \autoref{fig:CBOA}, the contributions due to zero-point
(photonic and vibrational) fluctuations only contribute negligibly to the change
in reaction rate in the Shin--Metiu model.

In general, a significant change of polarizability (either electronic or
vibrational, which can be comparable in some molecules~\cite{Bishop1990,
Bishop1998, Cammi1998}) from the equilibrium to the transition state
configuration could lead to similarly large effects as a change in the permanent
dipole moment, especially if the cavity frequency $\omega_c$ is relatively
large. However, it can be estimated that the vibrational contribution to the
zero-point energy shift is negligible for conditions typical for vibrational
strong coupling. To be precise, at resonance $\omega_c=\omega_v$, this reduces
to $-\Omega_R^2/(8\omega_v)$. Even for a relatively large vibropolariton Rabi
splitting of $\Omega_R \approx 0.2\omega_v$~\cite{Shalabney2015, Simpkins2015,
Dunkelberger2016}, this contribution is of the order of $\approx
10^{-2}\omega_v$, and thus small compared to typical barrier heights.

Finally, we note that the energy shifts above can be straightforwardly
generalized to the case of multiple cavity modes within second-order
perturbation theory. As can be easily verified, this simply leads to a sum over
modes $k$, giving a final energy shift
\begin{equation}\label{eq:deltaE_pert}
	\delta E(\bR) = -\sum_k \frac{\lambda_k^2}{2} \left( \mu_0^2(\bR)
	+ \frac{\omega_k}{2} \alpha_0(\bR) \right).
\end{equation}
This general expression, which is just the second-order energy correction due to
coupling to a set of cavity modes within the CBO, corresponds to the well-known
Casimir--Polder energy shift~\cite{Casimir1948}. The additional CBO
approximation, in which nonadiabatic transitions between electronic surfaces are
neglected, amounts to the approximation that the relevant cavity frequencies
$\omega_k$ are much smaller than the electronic excitation energies
$V_m(\bR)-V_0(\bR)$, such that only the (electronic) zero-frequency
polarizability $\alpha_0(\bR)$ appears in the second term. In contrast, the
first term depends only on the ground-state molecular permanent dipole moment
$\mu_0 = \langle 0|\hat\mu|0\rangle$, which does not involve electronically
excited states, and the CBOA thus does not amount to an additional
approximation. In \appref{app:nanoparticle_CP_vdW}, we demonstrate that, for the
case that the cavity can be approximated as a point-dipole (valid for a
sufficiently small nanoparticle), the perturbative energy shifts obtained here
correspond exactly to van der Waals forces~\cite{Stone2013}, with the first term
being the Debye force due to interaction between the permanent molecular and the
induced nanoparticle dipole, and the second term the London force due to
interaction between fluctuating dipoles. Under the point-dipole approximation,
the sum over cavity modes for the Debye force can again be rewritten in terms of
the zero-frequency polarizability of the nanoparticle.

\autoref{eq:deltaE_pert} is general for any kind of molecular reaction as long
as the light-matter coupling is not too large. It demonstrates that the most
relevant bare-molecule properties determining cavity-induced chemical reactions
in the ground state are the permanent dipole moment and polarizability close to
equilibrium, $\mu_0(\bR_0)$ and $\alpha_0(\bR_0)$, and transition
state, $\mu_0(\bR_\mathrm{TS})$ and $\alpha_0(\bR_\mathrm{TS})$,
configurations, and \emph{not} the transition dipole moment of the vibrational
excitation close to equilibrium, $\mu_\nu \propto \mu'_0(\bR_0)$, that
determines the Rabi splitting. In addition to changing reaction barriers, it
should be noted that the cavity-induced modification could potentially lead to a
plethora of diverse chemical modifications, such as a change of the relative
energy of different (meta-)stable ground-state configurations and thus a change
of the most stable configuration, or even the creation or disappearance of
stable configurations. Furthermore, depending on the particular properties of
the molecule, the cavity-induced change in the energy barriers can either lead
to suppression or acceleration of chemical reactions.

\section{Multimode cavity: nanoparticle-on-mirror}\label{sec:NPoM}

\begin{figure}
\includegraphics[width=\linewidth]{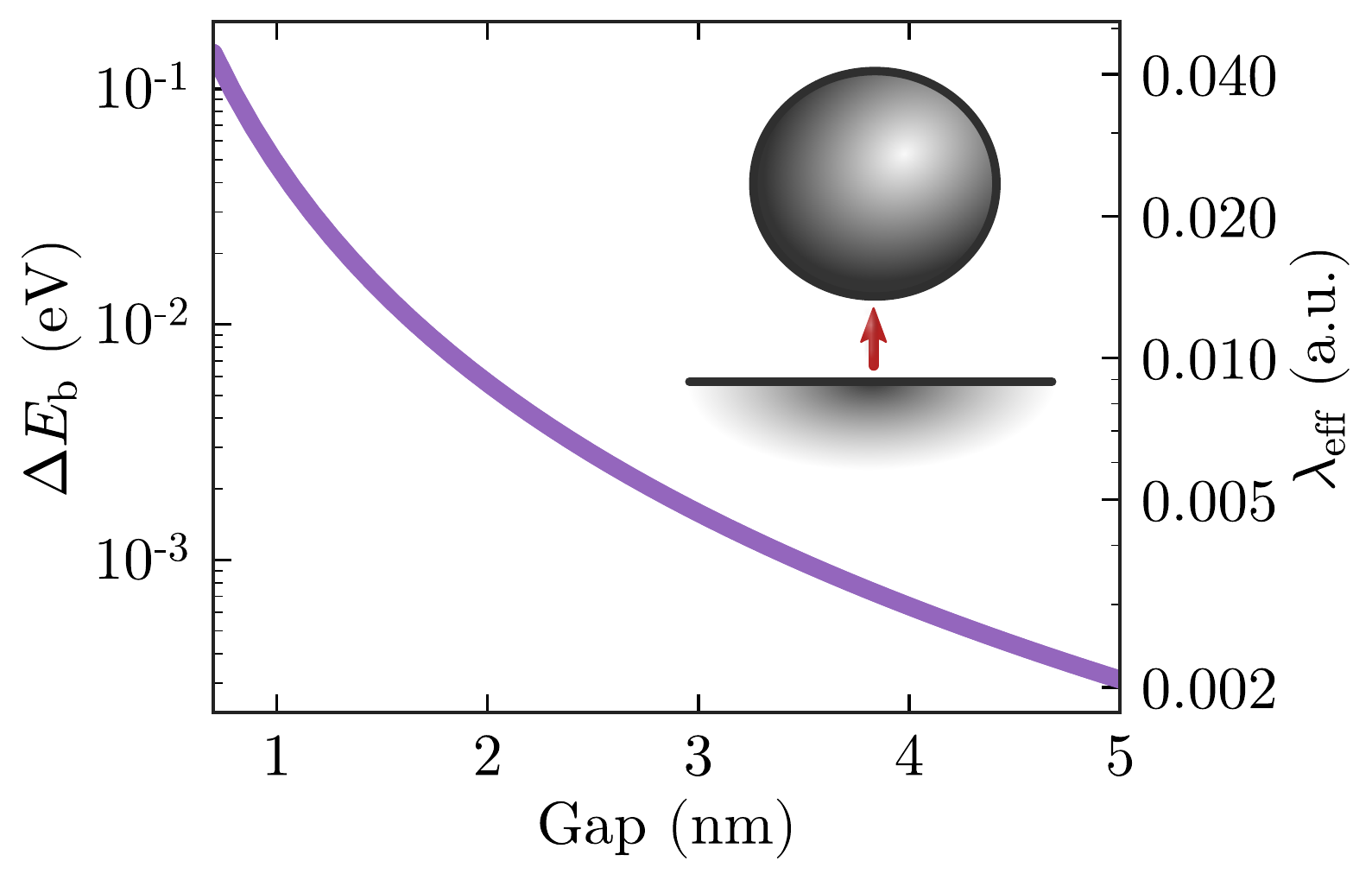}
\caption{Change of the energy barrier for the Shin-Metiu model inside a
nanoparticle-on-mirror cavity as a function of gap size. The right $y$-axis
shows the corresponding values of the effective single-molecule coupling
strength $\lambda_\mathrm{eff} = \sqrt{\sum_k\lambda_k^2}$. Inset: Illustration
of nanoparticle-on-mirror cavity geometry, with a single molecule placed in the
nanogap between a planar metallic surface and a small metallic nanoparticle of
radius $R=20$~nm.}
\label{fig:sphereonmirror}
\end{figure}

To demonstrate that the effects predicted above can be significant in realistic
systems, we treat a nanoparticle-on-mirror cavity with parameters taken from the
experiment in~\cite{Chikkaraddy2016}. This consists of a spherical metallic
nanoparticle (radius $R=20~$nm) separated by a small gap from a metallic plane,
see the inset of~\autoref{fig:sphereonmirror}. In this system, there is a series
of multipole modes coupled to the molecule~\cite{Li2016Transformation}, with
nontrivial behavior. Although several strategies can be employed to obtain the
quantized light modes in this system~\cite{Li2016Transformation, Urbieta2018},
we instead exploit that the dominant contribution we found above is due to
Debye-like electrostatic forces induced by the permanent molecular dipole, and
thus simply solve the electrostatic problem. To be precise, we calculate the
energy shift of a permanent dipole in this cavity as obtained by its interaction
with the field it induces in the cavity itself. Due to the simple involved
geometric shapes (a sphere and a plane), this can be achieved by the technique
of image charges and dipoles (see \appref{app:dipole_images} for details of the
calculation). We furthermore rely again on perturbation theory, i.e., we assume
that the molecular rearrangement due to its self-induced field is negligible.
Within this approximation, the energy shift we obtain from the purely
electrostatic calculation is equivalent to the term proportional to $\mu_0^2$ in
\autoref{eq:deltaE_pert}. The corresponding change $\Delta E_b$ in the height of
the energy barrier for the Shin-Metiu molecule is shown in
\autoref{fig:sphereonmirror} as a function of the gap size (as a point of
reference, the estimated gap size in~\cite{Chikkaraddy2016} is $0.9$~nm). We
find that the change in energy barrier can be significant, corresponding to a
change of the reaction rate by an order of magnitude or more
(cf.~\autoref{fig:pert_validity}). For comparison, in the figure we also show
the effective coupling strength $\lambda_\mathrm{eff} =
\sqrt{\sum_k\lambda_k^2}$ corresponding to each gap size. This value corresponds
to the coupling strength in a single-mode cavity that would give the same total
energy shift as obtained in this realistic multi-mode cavity. We note that we
have here treated a perfect spherical nanoparticle, and did not include
atomic-scale protrusions, which have been found to lead to even larger field
confinement due to atomic-scale lightning rod
effects~\cite{Benz2016,Urbieta2018,Lombardi2018}. For the experimental gap size
of $0.9$~nm, the effective coupling still becomes as large as
$\lambda_\mathrm{eff} \approx 0.031$~a.u., corresponding to $V_\mathrm{eff} =
4\pi/\lambda_\mathrm{eff}^2 \approx 1.9$~nm$^3$. This corresponds to a change in
the energy barrier of $\delta E_b \approx 0.07$~eV for the Shin-Metiu model
within second-order perturbation theory, which starts to break down at these
couplings, as we previously saw in~\autoref{fig:pert_validity}. This large
effective coupling demonstrates the importance of the multi-mode nature of these
cavities and the contribution of optically dark modes, as the ``bright'' nanogap
plasmon mode that is seen in scattering spectra has an estimated mode volume of
$\approx 40$~nm$^3$.

\section{Realistic molecule: 1,2-dichloroethane}\label{sec:dich}

\begin{figure*}
	\includegraphics[width=.95\linewidth]{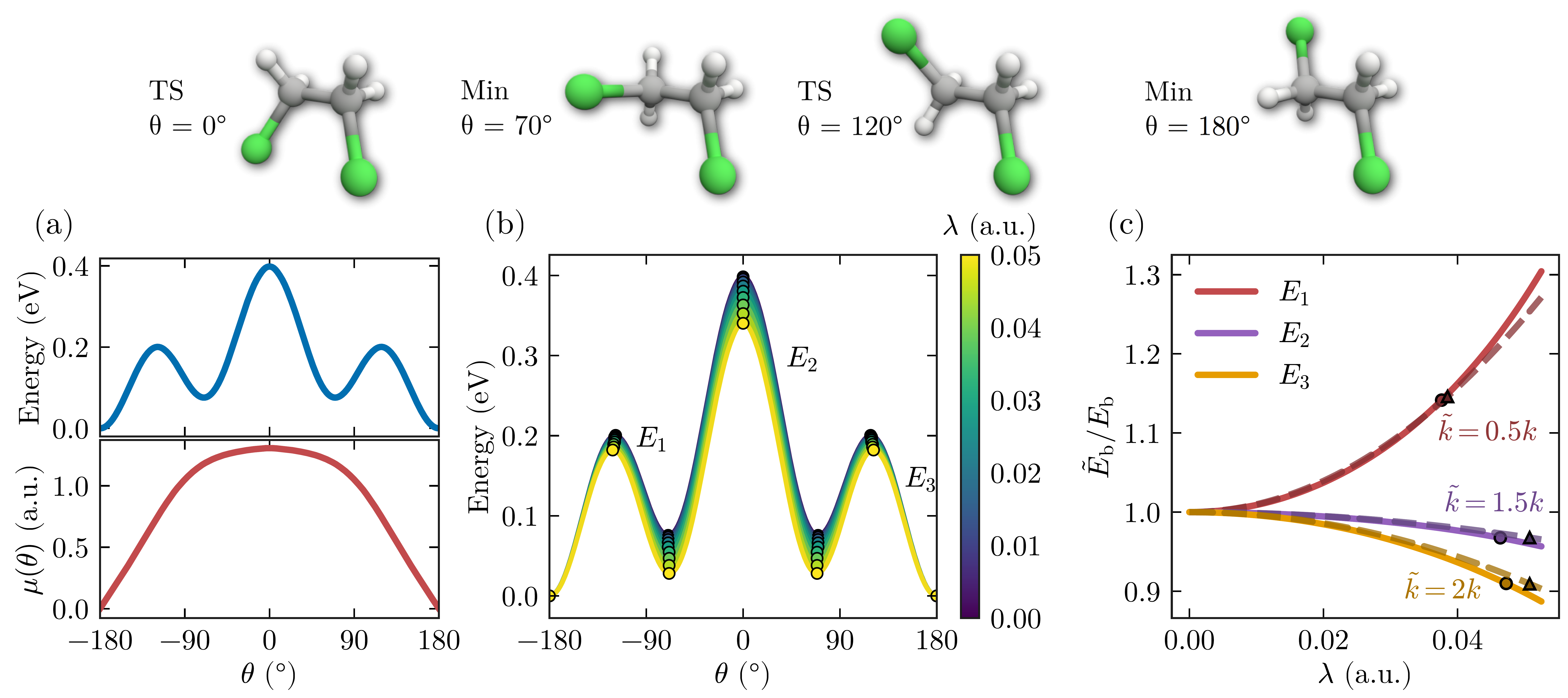}
	\caption{Top: Different configurations along the internal rotation of
	1,2-dichloroethane. (a) Energy landscape and dipole moment of the molecule.
	(b) Modified energy path for minimum $q$ for different coupling strengths.
	The energy barriers of the bare molecule are defined as $E_1 = V(120^\circ)
	- V(70^\circ)$, $E_2 = V(0^\circ) - V(70^\circ)$, and $E_3 = V(120^\circ) -
	V(180^\circ)$. (c) Relative modification of the energy barriers depending on
	the coupling strength for the full calculation (full lines, circles) and for
	perturbation theory (dashed lines, triangles). The marked points indicate
	relevant changes in the rate.}
	\label{fig:dichloroethane}
\end{figure*}

In the following, we apply the CBOA-TST theory to treat the internal rotation of
1,2-dichloroethane. In order to obtain the ground-state CBO surface under strong
light-matter coupling, we calculate the (ground and excited-state) bare-molecule
potential energy surfaces and permanent and transition dipole moments for a scan
along the rotation angle (defined as the Cl-C-C-Cl dihedral angle). For
simplicity, we here use the relaxed ground-state configuration of the bare
molecule for each rotation angle, i.e., we neglect cavity-induced changes in
degrees of freedom different from the internal rotation angle. The molecular
properties are obtained with density functional theory calculations with the
B3LYP~\cite{Becke1993} hybrid exchange-correlation functional and the
6-31$+$G(d) basis set. Excited states were computed with time-dependent density
functional theory within the Tamm-Dancoff approximation~\cite{Isborn2011}. All
calculations were performed with the TeraChem package~\cite{Ufimtsev2009,
Titov2013}.

The rather simple 1,2-dichloroethane molecule presents several characteristic
configurations along the rotation of the chlorine atoms around the axis defined
by the carbon-carbon bond (see top of \autoref{fig:dichloroethane}). It thus
constitutes an excellent model system to show several possible effects induced
by coupling to a cavity. In \autoref{fig:dichloroethane}(a) we present the
calculated ground state energy landscape and dipole moment, while some relevant
configurations are shown at the top. Analogously to the Shin--Metiu case, we
present the path of minimum energy along $q$ in \autoref{fig:dichloroethane}(b),
but here calculated within perturbation theory, \autoref{eq:V_at_qm}. We have
explicitly checked that the contribution due to London forces is negligible here
as well, and focus on the Debye-like contribution in the following. We see that
the most stable configuration ($\theta=180^\circ$) shows no change due to the
absence of a permanent dipole moment, while the most unstable one presents a
large energy shift. Therefore the different energy barriers of the system,
represented versus the coupling strength in \autoref{fig:dichloroethane}(c), are
altered significantly. Here we compare the energy barriers as predicted by
perturbation theory (dashed lines) with the ones from a full diagonalization of
the electronic Hamiltonian within the CBOA (full lines). In order to perform a
full calculation we have calculated the electronic potential energy surfaces and
the full dipole moment operator for a basis of $17$ electronic states. We also
indicate the points at which the coupling leads to important changes in the
relative rates calculated with TST, i.e., the coupling/energy at which we
achieve either suppression of $\tilde{k}/k = 0.5$ or enhancement of
$\tilde{k}/k=1.5$ or $2$. We see that in the case of perturbation theory
(triangles) the energy changes are slightly underestimated and thus larger
couplings are needed to reach the same rate change as in the full calculation
(circles).

As can be clearly seen, this still relatively simple molecule shows several
different kinds of phenomena. We see that the reaction rate out of the global
minimum at $\theta = 180^\circ$, corresponding to $E_3$, is increased. On the
other hand, $E_1$ increases and the local minimum situated at $\theta =
70^\circ$ is thus stabilized. \autoref{fig:dichloroethane}(b) suggests that this
effect could potentially become more dramatic for larger couplings than treated
here, as $\theta = 70^\circ$ could become the new global minimum of the system.
Finally, it is worth noting that the locations of the minima in energy also
change for larger couplings. This shift is most noticeable for the minimum at
$\theta = 70^\circ$, which transforms to $\tilde{\theta} \approx 68^\circ$ for
$\lambda = 0.05$~a.u..


\section{Resonance effects}\label{sec:resonance}

The results presented above predict a change in the ground-state reactivity that
is actually independent of the cavity photon frequency and in particular does
not rely on any resonance effects between the cavity mode and the vibrational
transitions of the molecule. Although the CBO PES can and does represent
vibro-polariton formation through normal-mode hybridization, as discussed above
and in \appref{app:normal_modes}, the subsequent TST used to predict
changes in chemical reaction rates is an inherently classical theory and does
not depend on the quantized frequencies of motion on the PES, and, as mentioned
above, neither on the transition dipole moment between vibrational levels
(determined by the derivative of the permanent dipole moment). While we have
shown that TST agrees almost perfectly with full quantum rate calculations,
where nuclear and photonic motion is quantized and polariton formation is thus
included, all calculations above have been performed for the resonant case
$\omega_c=\omega_\nu$.

\begin{figure}
	\includegraphics[width=\linewidth]{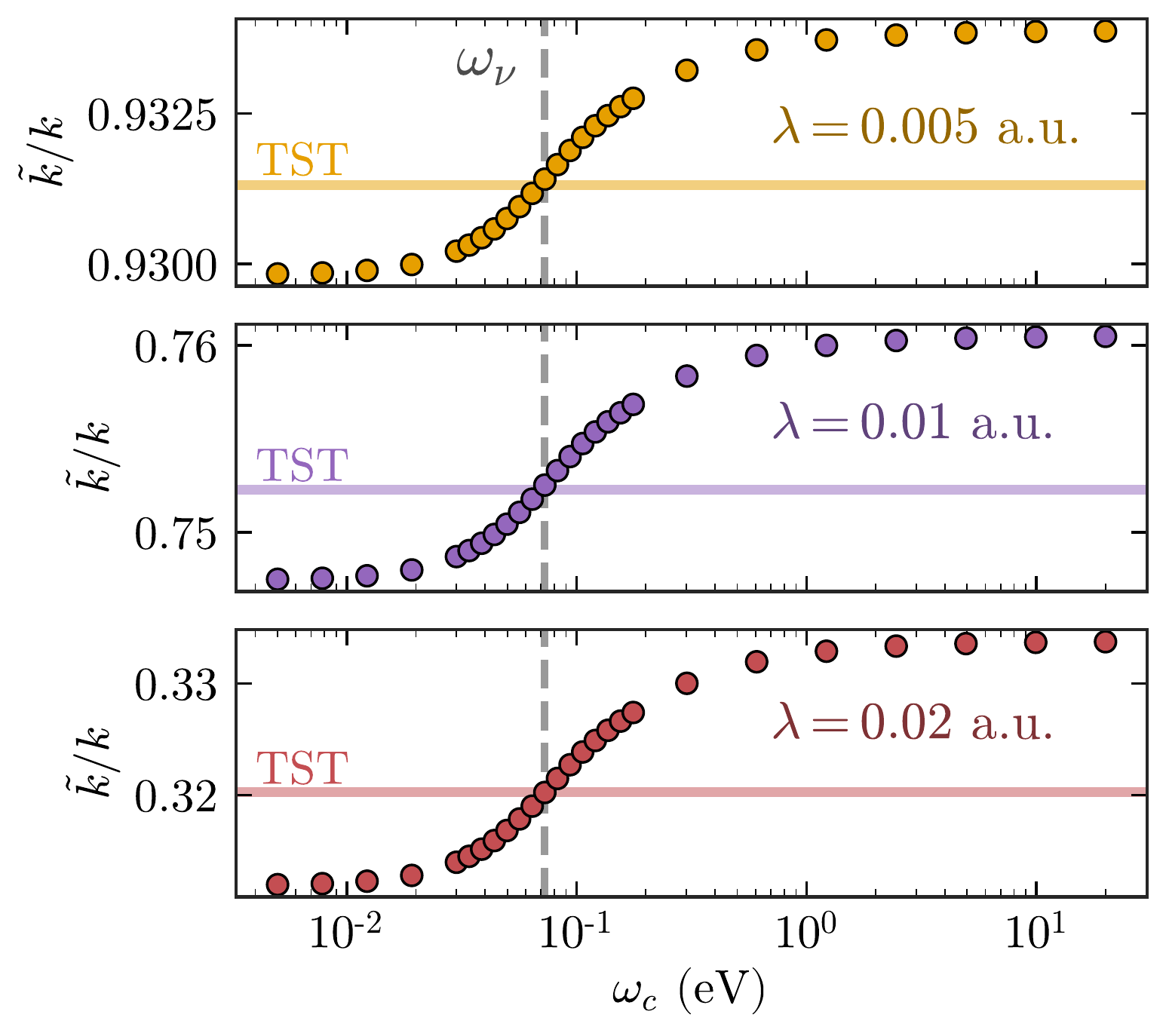}
	\caption{Ratio between on- ($\tilde{k}$) and off-cavity ($k$) rates vs the
	cavity frequency for three different values of the coupling strength. We
	increase the density of points close to the vibrational frequency of the
	molecule $\omega_\nu \approx 72.6$~meV in order to explore potential
	resonance effects.}
	\label{fig:TST}
\end{figure}

We thus investigate whether there is any resonance effect on chemical
ground-state reactivity by performing full quantum rate calculations for a wide
range of cavity frequencies within the Shin--Metiu model. In \autoref{fig:TST},
we represent the change $\tilde{k}/k$ in the calculated reaction rate of the
coupled system relative to the uncoupled molecule as a function of $\omega_c$,
for three different coupling strengths $\lambda$. Here, the values at
$\omega_c=\omega_\nu$ correspond to the results shown in \autoref{fig:CBOA}. We
observe that the cavity rates are essentially constant with the frequency, with
only a small modulation $(\tilde{k}(\omega_c\rightarrow \infty) -
\tilde{k}(\omega_c\rightarrow 0) \neq 0)$ that becomes more important for larger
couplings. For the cases represented in \autoref{fig:TST}, this goes from a
relative modulation of $0.4$\% for $\lambda=0.005$~a.u.\ to a $7$\% modulation
for $\lambda=0.02$~a.u.. However, no resonance effects are revealed close to the
vibrational frequency of the molecule, $\omega_\nu$. At the same time, the
vibrational frequency appears to be the relevant energy that separates the high-
and low-frequency limits for the rates, with TST working particularly well
exactly around that value. In the following, we show that both limits can be
understood by different additional adiabatic approximations.

In the high-frequency limit, $\omega_c \gg \omega_\nu$, the photonic degree of
freedom is fast compared to the vibrational one, and can thus be assumed to
instantaneously adapt to the current nuclear position $R$. This implies that the
photonic degree of freedom can be adiabatically separated (just like the
electronic ones), and nuclear motion takes place along an effective 1D-surface
determined by the local minimum in $q$, i.e., along the path sketched in
\autoref{fig:CBOA}(a), or, within lowest-order perturbation theory, along the
surface defined by \autoref{eq:V_at_qm}. Quantum rate calculations along this
effective 1D PES indeed reproduce the reaction rate in the high-frequency
limit perfectly (not shown). Furthermore, we note that in this limit, it becomes
convenient to directly group the photonic and electronic degrees of freedom to
obtain polaritonic PES~\cite{Galego2015,Feist2018} when performing the
Born--Oppenheimer approximation, as successfully used for electronic strong
coupling. In particular, this approach leads to exactly the same expression for
the effective ground-state PES~\cite{Galego2015}.

In the low-frequency limit, $\omega_c \ll \omega_\nu$, on the other hand, the
photonic motion is much slower than the vibrations and can also be adiabatically
separated. The photons are now too slow to adjust their configuration and $q$
can be assumed to stay constant during the reaction. The full quantum rate can
then be obtained by performing a thermal average of independent 1D quantum rate
calculations for each cut in $q$ of the two-dimensional surface
$\tilde{V}(R,q)$. Here, the (normalized) thermal weight at each $q$,
$\mathcal{P}(q)=\exp(-\langle E \rangle (q)/k_\mathrm{B}T)$, is calculated by
calculating the average thermal energy of the system $\langle E\rangle(q)$ for
constant $q$. Again, this approximation agrees perfectly with the full quantum
rate calculation for $\omega_c \to 0$ (not shown).

These results imply that, on the single-molecule level, the formation of
vibro-polaritons when $\omega_c\approx \omega_\nu$ is not actually required or
even relevant for the cavity-induced change in ground-state chemical structure
and reactivity. This fact can be appreciated by a simple intuitive argument:
vibrational strong coupling primarily occurs with the lowest vibrational
transitions close to the equilibrium configuration, while chemical reactions
that have to pass an appreciable barrier are typically determined by the
properties of the involved transition state, and the associated barrier height
relative to the ground-state configuration. In general, neither of these are
related to the properties of the lowest vibrational transitions (i.e., curvature
of the PES and derivative of the dipole moment at the minimum).

The absence of resonance effects can also be appreciated through the connection
to the well-known material-body-induced potentials obtained within perturbation
theory. For example, if the EM mode is well-approximated by a point-dipole mode,
the obtained energy shift in the CBO PES can be rewritten as a
van-der-Waals-like interaction between the permanent dipole moment of the
molecule and the dipole it induces in the nanoparticle. This corresponds to the
Debye force. In turn, the zero-point energy of the EM field reproduces the
London dispersive force due to vacuum fluctuations, and depends on the
polarizability of the molecule. For an arbitrary EM environment, this effect can
also be directly linked to Casimir--Polder forces~\cite{Casimir1948,
Sukenik1993}, which exactly correspond to the generalization of emitter-emitter
interactions to arbitrary material bodies (e.g., cavities). In particular,
within the perturbative regime, the applicability of Casimir--Polder approaches
could also be used to replace the explicit sum over modes $k$ by integrals
involving the EM Green's function~\cite{Buhmann2007Thesis, Scheel2008}, which is
readily available for arbitrary structures. This provides an additional argument
for the absence of resonance effects in our calculations, as (ground-state)
Casimir--Polder forces are well-known not to depend on resonances between light
and matter degrees of freedom.

While we do not explicitly treat the situation in recent experiments on the
modification of ground-state reactions by vibrational strong coupling (which
were found to depend strongly on resonance conditions~\cite{Thomas2016,
Hiura2018, Lather2018, Thomas2019}), we believe that our results indicate that
the resonance-dependent effects cannot be explained by a straightforward
modification of ground-state reaction energy barriers at thermal equilibrium, as
these would be captured by TST within the CBOA also in a many-mode,
many-molecule setting.


\section{Collective effects}\label{sec:collective}

We now turn to the description of collective effects, i.e., the case of multiple
molecules. For simplicity, we again restrict the discussion to a single cavity
EM mode. As discussed in \autoref{sec:rates}, the single-molecule effects we
have discussed up to now only become significant for coupling strengths $\lambda
= \sqrt{4\pi/V_\mathrm{eff}}$ corresponding to the smallest available plasmonic
cavities, which typically operate at optical frequencies. However, typical
experimental realizations of vibrational strong coupling are performed in
micrometer-size cavities filled with a large number of
molecules~\cite{Shalabney2015,George2015Liquid-Phase,Simpkins2015,Thomas2016}.
In this case, the per-molecule coupling $\lambda$ is so small that the
single-molecule effects discussed above are completely negligible. For strong
coupling and the associated formation of vibro-polaritons, the coherent response
of all molecules leads to a collective enhancement of the Rabi splitting
$\Omega_\mathrm{R,\mathrm{col}} = \sqrt{N} \Omega_\mathrm{R}$. However, as
we have seen that the cavity-induced modification of the single-molecule ground
state does not depend on the formation of polaritons, it is not a priori obvious
whether this collective enhancement of the Rabi splitting also translates to
cavity-induced collective modifications of the effective reaction barrier.

We thus repeat the analysis performed for the single-molecule case above for the
case of multiple molecules, working directly within the cavity Born--Oppenheimer
approach. We note that the arguments for its applicability for treating
ground-state chemical reactions translate straightforwardly from the single- to
the many-molecule case. For $N$ identical molecules, the CBO light-matter
interaction Hamiltonian becomes
\begin{multline}\label{eq:Hmanymol}
\hat{H}_{\mathrm{e}}^{(N)} = \frac{\omega_c^2}{2} q^2
+ \sum_i \left(\hat{H}_\mathrm{e}(\vecop{x}_i;\bR_i) + \omega_c q \blambda_i \cdot \vecopsym{\mu}(\vecop{x}_i;\bR_i) \right) \\
+ \sum_{i,j} \hat{H}_{\mathrm{dd}}(\vecop{x}_i,\vecop{x}_j;\bR_i,\bR_j),
\end{multline}
where $\hat{H}_{\mathrm{dd}}$ accounts for direct intermolecular (dipole-dipole)
interactions. We stress that we again assume that only a single cavity mode is
significantly coupled to the molecules. The cavity-mediated dipole-dipole
interaction is thus fully contained within the light-matter coupling term, and
$\hat{H}_{\mathrm{dd}}$ corresponds to the free-space
expression~\cite{DeBernardis2018Cavity}. In the following discussion, we will
again use lowest-order perturbation theory to obtain analytical insight. The
cavity-molecule and dipole-dipole interaction terms are then independent
additive corrections. We first focus on the cavity-induced effects, and will
discuss the influence of direct dipole-dipole interactions later, in particular
when studying a prototype implementation: A nanosphere surrounded by a
collection of molecules. For simplicity of notation, we again use scalar
quantities to indicate the component of the dipole along the field direction,
but keep the index $\epsilon$ to make this explicit, i.e., $\blambda_i =
\lambda_i \boldsymbol{\epsilon}_i$ and $\boldsymbol{\epsilon}_i \cdot
\bmu(\bR_i) = \mu_\epsilon(\bR_i)$, so that we can rewrite the interaction term
of the Hamiltonian as $\omega_c q \sum_i^N \lambda_i
\hat{\mu}_\epsilon(\vecop{x}_i;\bR_i)$. The full Hamiltonian now corresponds to
a many-body problem even for simple model molecules. Within second-order
perturbation theory, the new (many-molecule) ground-state PES is
\begin{multline}
\tilde{V}_0^{(N)}(\bR_t,q) = \sum_i V_0(\bR_i) + \frac{\omega_c^2}{2}q^2 \\
+ \omega_c q \sum_i \lambda_i \mu_{0,\epsilon}(\bR_i) - \frac{\omega_c^2}{2} q^2 \sum_i \lambda_i^2 \alpha_{0,\epsilon\epsilon}(\bR_i),
\end{multline}
where $\bR_t=(\bR_1,\bR_2,\dots,\bR_N)$ collects the nuclear configurations of
all the molecules. With this result, we can again apply the corresponding
conditions for finding critical points in order to analytically find the minimum
along $q$ and the corresponding total energy of the hybrid system up to second
order in $\lambda_i$,
\begin{equation}\label{eq:VN_at_qm}
\tilde{V}_0^{(N)}(\bR_t,q_\mathrm{m}) = \sum_i V_0(\bR_i) - \frac{1}{2}\left(\sum_i \lambda_i \mu_{0,\epsilon}(\bR_i)\right)^2.
\end{equation}
It can be seen that the cavity-induced shift depends on the square of the sum of
the (coupling-weighted) permanent dipole moments of the molecules, not on the
sum of their squares. Assuming perfect alignment and identical configurations
for all molecules, this gives an energy shift $-N^2 \bar\lambda^2
|\bmu_0(\bR)|^2$, where $\bar\lambda = \frac1N \sum_i \lambda_i$ is the average
coupling. The per-molecule energy shift is then linear in $N$, indicating
collective enhancement of the molecule-cavity interaction. In contrast, the
London-force-like change in zero-point energy due to the modification of the
effective cavity frequency is additive,
\begin{equation}
	\omega_\mathrm{eff} = \omega_c - \frac{\omega_c}{2} \sum_i \lambda_i^2 \alpha_{0,\epsilon\epsilon}(\bR_i) + \mathcal{O}(\lambda_i^4),
\end{equation}
with a total zero-point energy shift $\frac12 (\omega_\mathrm{eff} - \omega_c)$
proportional to $N$, and shows no collective enhancement for single-molecule
reactions. It is interesting to note that the connection between
polarizability and the dielectric function of a material through the
Clausius-Mossotti relation suggests that this energy shift is equivalent to the
change of mode frequency due to the refractive index of the collection of
molecules. The shift in cavity mode frequencies due to refractive index changes
after chemical reactions is exactly the effect used in experiments to monitor
reaction rates under vibrational strong coupling~\cite{Thomas2016, Thomas2019,
Hiura2018}. We also mention that at higher levels of perturbation theory,
cavity-mediated contributions analogous to the Axilrod-Teller potential, i.e.,
van-der-Waals interactions between three emitters, appear in the intermolecular
potential~\cite{Axilrod1943,Scheel2008}.

Based on \autoref{eq:VN_at_qm}, we can analyze the effect of the cavity on the
reaction rate of a single molecule within the ensemble. This is determined by
the energy difference between minimum-energy and transition-state configurations
of that molecule, with the other molecules fixed in a stable configuration (here
chosen to be the minimum for all of them). For simplicity, we assume that the
critical configurations $\bR_\mathrm{Min}$ and $\bR_\mathrm{TS}$ of the coupled
system are equal to the uncoupled ones (as we have seen above, the shifts are
generally small). We can then directly express the change in the energy barrier
of the moving molecule (chosen to be molecule $i=1$ here) as
\begin{multline}\label{eq:collective_Eb_0}
	\tilde{E}_\mathrm{b} = E_\mathrm{b} - \frac{\lambda_1^2}{2}
	\left(\mu_{0,\epsilon}^2(\bR_\mathrm{1,TS}) - \mu_{0,\epsilon}^2(\bR_\mathrm{1,Min}) \right) - \\
	\lambda_1 \left(\sum_{i=2}^N \lambda_i \mu_{0,\epsilon}(\bR_\mathrm{i,Min})\right)
	\left(\mu_{0,\epsilon}(\bR_\mathrm{1,TS}) - \mu_{0,\epsilon}(\bR_\mathrm{1,Min}) \right).
\end{multline}
This expression can be straightforwardly interpreted, with the first part
corresponding to the Debye-like interaction of molecule $1$ itself with the
cavity, and the second part corresponding to the cavity-mediated interaction of
molecule $1$ with all other molecules (which itself can be understood as the sum
of two equal contributions, the interaction of the moving molecule with the
cavity field induced by all other molecules, as well as the interaction of all
other molecules with the cavity field induced by the molecule). Within
perturbation theory, this Debye-like energy shift is again equivalent to the
electrostatic energy, in this case that of a collection of permanent dipoles
interacting with the cavity, i.e., a material structure. This makes the
connection to electric field catalysis~\cite{Fried2017} even more direct, with
the difference that the electric field is not generated by applying an external
voltage, but represents the cavity-enhanced field of all the other molecules.
The fact that the main contribution is just the electrostatic energy shift also
demonstrates the equivalence of our results to the approach of taking into
account non-resonant effects through cavity-modified dipole-dipole and
dipole-self interactions~\cite{DeBernardis2018Cavity}.

To treat the dependence on molecular orientations explicitly, we define the
alignment angle $\theta_i$ for each molecule through $\mu_{0,\epsilon}(\bR_i) =
|\bmu_0(\bR_i)| \cos\theta_i$. Inserting this in \autoref{eq:collective_Eb_0}, 
we obtain
\begin{multline}\label{eq:collective_Eb}
	\tilde{E}_\mathrm{b} = E_\mathrm{b} - \frac{\lambda_1^2}{2}
	\left(\mu_{0,\epsilon}^2(\bR_\mathrm{1,TS}) - \mu_{0,\epsilon}^2(\bR_\mathrm{1,Min}) \right) - \\
	N' \bar\lambda^2 \langle \cos\theta \rangle' |\bmu_{0}(\bR_\mathrm{Min})| \times \\
	\lambda_{r,1} \left(\mu_{0,\epsilon}(\bR_\mathrm{1,TS}) - \mu_{0,\epsilon}(\bR_\mathrm{1,Min}) \right),
\end{multline}
where $\lambda_{r,i} = \lambda_i / \bar\lambda$ is the relative coupling of
molecule $i$, $\langle\cos\theta\rangle = \frac1N \sum_i \lambda_{r,i}
\cos\theta_i$ is the coupling-weighted average orientation angle, and primed
quantities indicate that only molecules $2$ to $N$ are taken into account (for
$N\gg1$, they can be replaced by unprimed quantities). We obtain a term
proportional to the number of molecules $N$, i.e., there is a collective effect
on the single-molecule energy barrier that is reminiscent of the collective Rabi
splitting, $N \lambda^2 \propto \Omega_{R,\mathrm{col}}^{2}$. Note that the
collective change of the energy barrier still depends on the molecule having a
different permanent dipole moment in the transition and minimum configuration.
Furthermore, it requires the molecules not participating in the reaction to have
a non-zero permanent dipole moment and an average global alignment, such that
$\langle\cos\theta\rangle\not=0$. This could be achieved by fixing the molecular
orientation by, e.g., growing self-assembled monolayers~\cite{Nicosia2013} or
using DNA origami~\cite{Acuna2012, Chikkaraddy2018}, or for molecules that can
be grown in a crystalline phase, such as anthracene~\cite{Kena-Cohen2008}
(although polar molecules tend not to grow into crystals with a global
alignment~\cite{Hulliger2012}). Another strategy to achieve alignment under
strong coupling that has been successfully used experimentally is to align
molecular liquid crystals through an applied static field~\cite{Hertzog2017}.
However, for general disordered media such as polymers or molecules flowing in
liquid phase~\cite{Thomas2016, George2015Liquid-Phase}, the angular distribution
is typically isotropic, leading to $\langle \cos\theta \rangle \approx 0$. In
that case, our theory predicts that no collective effect on reactivity should be
observed unless the cavity itself induces molecular orientation (see below). We
note for completeness that the collective Rabi splitting depends on the average
of the squared $z$-component of the transition dipole moments, i.e.,
$\langle\cos^2\theta\rangle$, which is nonzero unless all molecules are aligned
perpendicular to the electric field of the cavity mode, and equal to $1/3$ for
isotropic molecules.

\begin{figure*}
	\includegraphics[width=\linewidth]{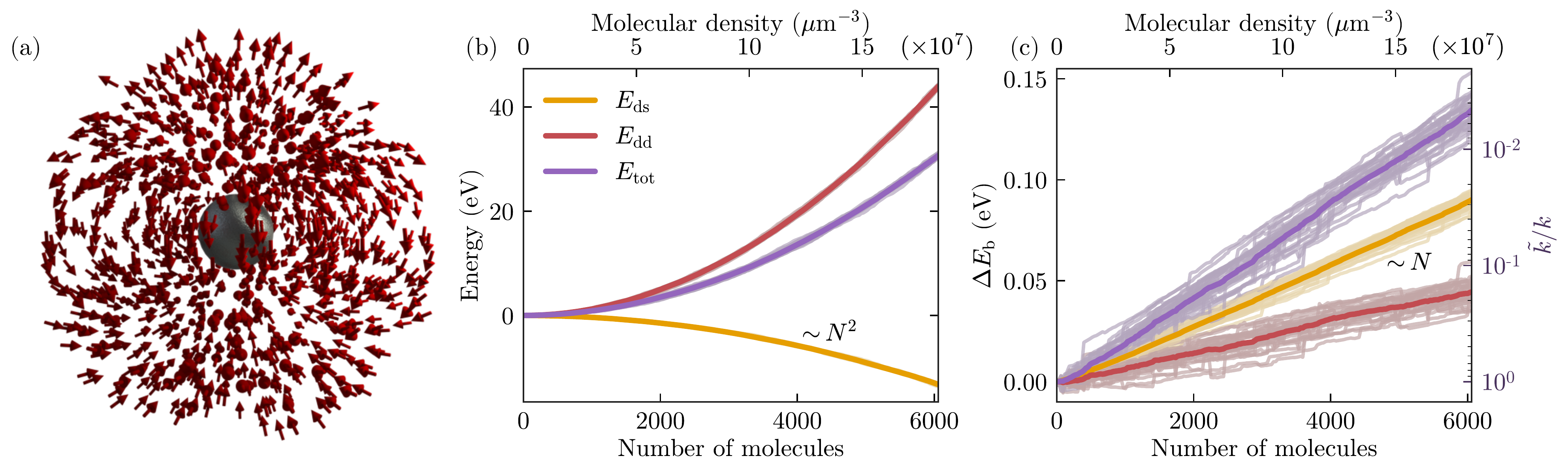}
	\caption{(a) Sketch of the model system of a collection of molecules
	distributed around a metal nanosphere with a diameter of $8$~nm. The
	molecules are placed randomly at distances from $1$~nm to $16$~nm to the
	surface of the sphere, with the (permanent) dipoles aligned along the
	direction of the field of the sphere's $z$-oriented dipole mode. (b) Energy
	due to the dipole-sphere ($E_{\mathrm{ds}}$) and dipole-dipole
	($E_{\mathrm{dd}}$) interactions in the system within perturbation theory as
	a function of number of molecules $N$, as well as their sum
	($E_{\mathrm{tot}}$). (c) Change in energy barrier and corresponding change
	in reaction rate at room temperature for the most strongly coupled molecule,
	also resolved into contributions from dipole-sphere and dipole-dipole
	interactions. In both panels (b) and (c), the slightly transparent lines
	correspond to different random realizations of the system, with the averages
	in solid lines.}
	\label{fig:many_mol_sphere}
\end{figure*}

In order to test the strength of the collective effect in realistic situations,
and to compare it with the effect of direct (free-space) dipole-dipole
interactions, we now treat a specific configuration, as depicted in
\autoref{fig:many_mol_sphere}(a): A nanocavity represented by a metallic sphere
of diameter $d=8$~nm, surrounded by a collection of Shin--Metiu ``molecules'',
located at distances from $1$~nm to $16$~nm from the sphere. We place a
collection of up to $N=6000$ molecules at random positions within that volume,
imposing a minimum distance of $1.5$~nm between the molecules. A metal sphere
with a Drude dielectric function (or a dielectric sphere with a single
resonance, such as a phonon mode) can be approximated as a cavity with only
three modes, the dipolar localized surface plasmon resonances aligned along $x$,
$y$, and $z$ (see \appref{app:sphere_quantization} for details). Higher order
multipole modes only couple significantly to emitters that are very close to the
surface. We first assume all molecules to be aligned perfectly with the electric
field of the $z$-oriented dipolar mode of the sphere. In this configuration, the
sum over $x$- and $y$-oriented fields at the origin cancels out for large $N$.
For these directions, there is thus no Debye-like collective effect, and we can
restrict our attention to just a single mode of the sphere (the $z$-oriented
dipole mode)\footnote{We have additionally checked explicitly that solving the
full electrostatic problem, i.e., including all modes of the sphere by using the
method of image dipoles, gives very similar results to the ones presented here}.
As mentioned above, within perturbation theory, where the Debye-force like
contribution can be understood within a fully electrostatic picture, it is
straightforward to include the direct (free-space)
permanent-dipole--permanent-dipole interaction, as it is simply a further
additive electrostatic contribution. In \autoref{fig:many_mol_sphere}(b), we
show the total electrostatic energy of the system, as well as the relative
contributions due to molecule-sphere and direct molecule-molecule interactions,
as a function of $N$. For the configuration considered here, for which we have
not performed any optimization of total energy, the dipole-dipole interactions
give a positive contribution to the total energy that is significantly larger
than the collective dipole-sphere interaction. The relative strength of
dipole-dipole and dipole-sphere interactions depends on the details of the
configuration, and we have checked that, e.g., it is also possible to maintain
the same collective interaction while obtaining an overall negative contribution
from dipole-dipole interactions by not choosing random positions as we did for
simplicity. 

In contrast to the total energy, the change in energy barrier predicted
by~\autoref{eq:collective_Eb_0} for the most strongly coupled molecule of the
ensemble is dominated by the (collective) sphere-dipole interactions, as shown
in \autoref{fig:many_mol_sphere}(c). The barrier height indeed increases
approximately linearly with $N$, with changes of up to $\approx 0.09$~eV due to
the cavity-mediated interaction, and an associated suppression of the reaction
rate by a factor of $\approx\!30$ at room temperature. In the geometry treated
here, the energy shift of the target molecule due to dipole-dipole interactions
with the other molecules also increases linearly with $N$, as the molecular
dipoles combine to all act in the same direction at the sphere location, with an
effect that is roughly half of the cavity-mediated interaction. As mentioned
above, the details depend strongly on the configuration and cavity properties,
and in particular, it is also possible to choose configurations where the direct
dipole-dipole interactions dominate. While a more exhaustive treatment is beyond
the scope of this article, we mention that in initial explorations, we did not
find any simple configuration where the cavity-mediated interactions were
significantly larger than direct dipole-dipole interactions.

\begin{figure}
	\includegraphics[width=\linewidth]{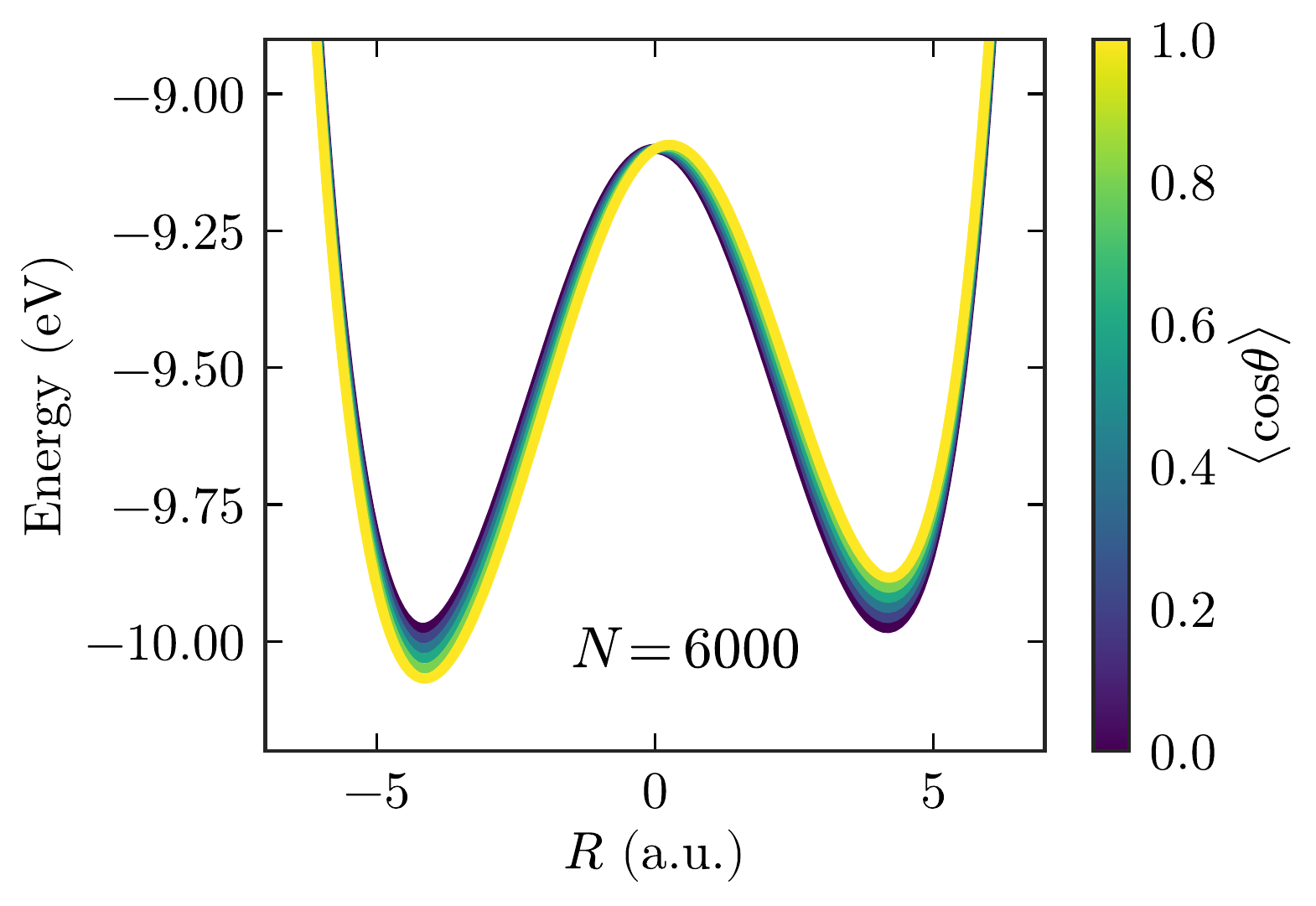}
	\caption{Alignment dependence of the cavity Born--Oppenheimer PES along the
	photonic minimum path $q_m$ for the molecule in
	\autoref{fig:many_mol_sphere}(c), with all other molecules fixed to the
	equilibrium position $R_{\mathrm{min}}$.}
	\label{fig:PES_sphere_alignment}
\end{figure}

While the barrier height increases here, the effect we predict can also lead to
a decrease, for example in the case that the transition-state dipole moment is
larger than in the minimum configuration, cf.~\autoref{eq:collective_Eb}. This
would be expected, e.g., in dissociation reactions in which the molecule splits
into two partially charged fragments, and is also seen for the backreaction from
the right to left minimum in the Shin--Metiu model for the case that all other
molecules are in the leftmost minimum (see \autoref{fig:PES_sphere_alignment}).

For comparison, \autoref{fig:PES_sphere_alignment} shows the effect of average
alignment for the sphere-molecule system considered above, for the case of
$N=6000$ molecules corresponding to a molecular density of $\approx
2\cdot10^8~\mu$m$^{-3}$. It displays the CBO PES within second-order
perturbation theory as a function of $R_1$, with all other molecules fixed in
the minimum configuration, and along the photonic minimum $q=q_m$. For
$\langle\cos\theta\rangle=1$, this demonstrates that the collective cavity
effect on the surface is significant, with the position of the critical points
shifting compared to the bare molecule. For the Shin--Metiu model studied here,
the barrier height is actually increased compared to the approximate prediction
\autoref{eq:collective_Eb_0}, which does not take into account these shifts. In
contrast, when there is no average orientation, $\langle\cos\theta\rangle=0$,
the effect on the surface is minimal and is reduced to the single-molecule
result.

The single-molecule energy shifts we predict for perfect alignment can be
significant. This implies that the molecules, if they are free to rotate in
place, could lower their energy by aligning with the electric field of the
cavity mode, which could possibly lead to self-organization (for the example
system above, this also requires breaking of the overall spherical symmetry).
The details of this effect depend on the precise setup, such as the cavity
material and shape, molecular and solvent properties, etc., and would require a
more complete treatment taking thermodynamical effects and free energy into
account~\cite{Chipot2007, Mendieta-Moreno2016}, which is beyond the scope of the
current work. However, we mention that it has recently been shown that strong
coupling and the associated formation of polaritons itself could lead to
alignment due to the associated decrease of the lower polariton energy, provided
that a significant fraction of molecules are excited to lower polariton
states~\cite{Cortese2017,Keeling2018}. Although thermal excitation can be
efficient for vibrational strong coupling due to the relatively low energies of
vibro-polaritons, on the order of a few times the thermal energy $k_B T$, it
should be noted that the arguments in~\cite{Cortese2017,Keeling2018} do not
directly translate to thermal-equilibrium situations. In that case, a change in
state energy due to improved orientation also leads to a change in population,
with the average energy per degree of freedom staying constant and thus no net
energy gain.

Finally, we mention that in contrast to the single-molecule case, the
generalization of the above arguments to many cavity modes is not
straightforward, and the results are thus not directly applicable to, e.g.,
Fabry-Perot cavities with a continuum of modes following a dispersion relation
as a function of the in-plane wave vector, as employed in existing
experiments~\cite{Thomas2016,Hiura2018,Lather2018,Thomas2019}. Our results
indicate that solving the electrostatic problem (where all modes are implicitly
taken into account) should predict the changes in energy barriers, but, e.g.,
the scaling with number of molecules is not immediately obvious, and as
mentioned above, statistical effects should be treated more carefully. Only for
the special case that all modes have the same electric field distribution (e.g.,
different dipolar resonances of a small nanoparticle), the sum over modes can be
performed straightforwardly.


\section{Conclusions}
To summarize, we have analyzed modifications of ground-state chemical reactivity
in hybrid cavity-molecule systems, motivated by experimental results showing
this for vibrational strong coupling~\cite{Thomas2016,Thomas2019}. By treating a
simple model system, the Shin--Metiu model, we were able to show through full
quantum rate calculations on the single-molecule level that ground-state
thermally driven reaction rates can indeed be significantly modified under
strong light-matter coupling. We then demonstrated that this change can be
interpreted through classical transition state theory, i.e., by the change in
the height of an effective energy barrier (or activation energy) by working
within the cavity Born--Oppenheimer approximation. In this approximation, the
cavity photon is formally treated like a nucleus, such that ground-state
reactions can be represented through motion on a PES with a single additional
nuclear-like degree of freedom. The use of perturbation theory leads to simple
analytic expressions relating the effective barrier heights to purely
ground-state molecular properties, namely the uncoupled ground-state PES, dipole
moment, and polarizability of the molecule. We showed that within second-order
perturbation theory, the energy shifts determining the barrier height on the CBO
PES can be directly related to well-known intermolecular forces, i.e., the Debye
and London forces, and more generally to Casimir--Polder interactions.

We stress that while perturbation theory allows us to make connections to
well-known results, our approach generalizes Casimir--Polder forces beyond the
perturbative regime and applies for any coupling strength. Additionally, we have
shown explicitly that the emergence of vibrational strong coupling does not
affect the validity of the derived expressions for the effective energy
barriers. At the same time, the CBOA provides a straightforward way to connect
to well-known theories of chemical reactivity. The fact that the energy shifts
obtained here become appreciable for realistic nanocavities with strongly
sub-wavelength field confinement and thus sufficiently large $\lambda$
demonstrates that the (generalized) van-der-Waals forces due to the interaction
of the molecular dipole with the polarization it induces in the cavity can
become strong enough to lead to significant changes in chemical reactivity.

We also note that in the context of Casimir--Polder forces, it is well-known that
for sub-wavelength separations between emitters and material systems, it is
sufficient to work within the quasistatic approximation, in which only the
longitudinal electromagnetic Green's function contributes and the interaction
does not depend on whether the Power-Zienau-Woolley transformation has been
performed or not. In this context, it is also well-known how to go beyond the
quasistatic approximation, and the contribution from longitudinal and
transversal fields (including the $\vecop{A}{}^2$ term and \emph{all} EM field
modes) is naturally included within the Green's
function~\cite{Buhmann2007Thesis}.

We demonstrated the applicability of our approach for a realistic multi-mode
cavity, a nanoparticle-on-mirror setup~\cite{Chikkaraddy2016}, and found that
the effective single-molecule coupling strength in this case becomes significant
(corresponding to a mode volume of $\approx 2$~nm$^3$) even though the mode
volume of the main optically active mode is significantly larger ($\approx
40$~nm$^3$). We furthermore applied our theory to a real molecule,
1,2-dichloroethane, and showed that reaction rates can be both suppressed and
enhanced depending on the relative value of the molecular dipole moment at the
critical configurations (local minima and saddle points of the PES). A cavity
could thus serve as a catalyst or as an inhibitor of a ground-state reaction,
and could even alter the global equilibrium configuration of the molecule, all
without any kind of external energy input, with all reactions simply driven by
thermal fluctuations. This represents a potential way to efficiently optimize
the desired yield of a molecular reaction.

We then found that on the single-molecule level, the effects discussed above do
not rely on any particular relation between the cavity photon frequency
$\omega_c$ and the vibrational transitions in the molecule $\omega_\nu$, and
thus in particular not on the formation of polaritons (hybrid light-matter
states). This is consistent with the interpretation of the energy shifts as
generalizations of Casimir--Polder interactions beyond the perturbative regime.
We also showed that the small modulation of the reaction rate as a function of
$\omega_c$ that is observed numerically can be understood by simple adiabatic
approximations, and again is not related to polariton formation.

For the case of many-molecule strong coupling, where the single-molecule
coupling $\lambda$ is typically so small that the single-molecule effects
described above are negligible, we demonstrated that the PES and reaction
barriers can be significantly modified by collective effects provided that the
permanent dipole moments of the molecules are oriented with respect to the
cavity mode field, such that they induce an overall static electric field.
However, it should also be noted that similar effects could be achieved by
direct dipole-dipole interactions if one manages to align all molecules such as
to create a strong field at the position of a single molecule. An interesting
open question is whether the cavity-mediated interactions could induce alignment
in materials that do not show this in the absence of the cavity, or if direct
dipole-dipole interactions would prevent this.

Finally, it should be noted that we have throughout assumed that the whole
system is in thermal equilibrium, i.e., that the effective temperature is
identical both for the molecules and the cavity EM mode. This implies that
system-bath interactions do not have to be explicitly modelled, as the system
can simply be assumed to be at a given temperature (as explicitly included in
the quantum rate calculations and TST). This assumption would break down if the
internal vibrational temperature of the molecules is different from the
temperature of the thermal radiation bath that the cavity is coupled to. In that
case, the effective temperature of the system could potentially become an
average of the internal and external bath temperatures. In particular, the
effective temperature relevant for a given reaction could depend on whether
vibrational motion along that reaction coordinate is hybridized with the cavity
mode, such that the external black-body radiation bath would conceivably couple
more efficiently to that mode than to others. Such effects have been studied for
Casimir--Polder forces, where resonant contributions that exactly cancel at
thermal equilibrium can become important in non-equilibrium
situations~\cite{Buhmann2008,Ellingsen2010}, and possibly give rise to
additional collective effects~\cite{Sinha2018}.

Our work demonstrates the possibility of modifying ground-state chemical
reactions and molecular properties in hybrid cavity-molecule systems without an
external input of energy. We believe that the theory presented here lays the
groundwork for a profound understanding of this novel cavity effect and could be
used to predict experimentally available chemical modifications.


\begin{acknowledgments}
We are grateful to M.~Ruggenthaler, S.~De~Liberato, and P.~Rabl for helpful
discussions. This work has been funded by the European Research Council
(ERC-2016-STG-714870) and the Spanish MINECO under contract MAT2014-53432-C5-5-R
and the ``Mar\'ia de Maeztu'' programme for Units of Excellence in R\&D
(MDM-2014-0377), as well as through a Ramón y Cajal grant (JF).
\end{acknowledgments}

\begin{appendix}

\section{Quantized modes of spherical nanoparticles}\label{app:sphere_quantization}
We here show that for two general models for the dielectric function, a
spherical nanoparticle in vacuum within the quasistatic approximation can be
approximated as a three-mode cavity~\cite{Waks2010, Gonzalez-Ballestero2015},
with the modes corresponding to three degenerate dipole modes. We treat the
sphere as a point dipole, equivalent to neglecting the short-range higher-order
multipole modes that only couple to emitters very close to the
surface~\cite{Anger2006,Delga2014}. The direction-independent polarizability of
the sphere is then given by~\cite{deVvanCLag1998}
\begin{equation}
	\alpha_S(\omega) = a^3 \frac{\epsilon(\omega) - 1}{\epsilon(\omega) + 2},
\end{equation}
where $a$ is the radius of the sphere. For a metallic Drude model dielectric
function without losses, $\epsilon_m(\omega) = 1 - \omega_p^2/\omega^2$, this
can be rewritten as
\begin{equation}
	\alpha_S(\omega) = \frac{a^3 \omega_0^2}{\omega_0^2 - \omega^2},
\end{equation}
where $\omega_0 = \omega_p/\sqrt{3}$. This is identical to the polarizability of
a single-mode quantum oscillator at frequency $\omega_0$ with transition dipole
moment $\mu_{eg} = \sqrt{\omega_0 a^3/2}$~\cite{Bonin1997},
\begin{equation}\label{eq:quant_polarizability}
	\alpha_q(\omega) = \mu_{eg}^2 \left(\frac{1}{\omega_0 - \omega} + \frac{1}{\omega_0 + \omega} \right)
	= \frac{a^3 \omega_0^2}{\omega_0^2 - \omega^2}.
\end{equation}
Here, spherical symmetry implies that there are three degenerate quantum
oscillators, corresponding to the quantized localized surface plasmon resonances
in this case, directed along three orthogonal axes (e.g., $x$, $y$, and $z$). If
the dielectric function is instead given by Lorentzian function representing a
material resonance (e.g., a phonon mode) at frequency $\omega_{\mathrm{ph}}$ and
with resonator strength characterized by $\omega_f$, i.e., $\epsilon(\omega) = 1
+ \frac{\omega_f^2}{\omega_{\mathrm{ph}}^2 - \omega^2}$, we again get the same
polarizability by using $\omega_0^2 = \omega_{\mathrm{ph}}^2 +
\frac{\omega_f^2}{3}$ and $\mu_{eg} = \omega_f\sqrt{\frac{a^3}{6\omega_0}}$,
with the quantized mode now corresponding to a localized surface phonon
polariton resonance. We have thus found that these simple models can be
quantized by considering just a single or few cavity modes.

\section{Normal modes in cavity Born--Oppenheimer}\label{app:normal_modes}

In this section, we demonstrate how vibro-polariton formation can be observed
within the cavity Born--Oppenheimer approximation~\cite{Flick2017Atoms}. We note
that this derivation is essentially identical to that performed
in~\cite{Shalabney2015}, although it was not based on the CBOA there. Within
this approximation, discussed in detail in \autoref{sec:CBOA}, the photonic
degree of freedom is described by a continuous parameter $q$, proportional to
the electric displacement field, and both nuclear and photonic motion takes
places on an electronic potential energy surface parametric in $\bR$ and
$q$. Hybridization of photonic and vibrational excitations is thus not directly
observed by inspecting the PES, but requires calculating the coupled
nuclear-photonic eigenstates determined by the PES\@. This is most easily
achieved close to a local minimum, where the surface can be locally approximated
through coupled harmonic oscillator potentials, and direct diagonalization of
the Hessian gives the polariton eigenstates. Close to the minimum and only
treating a single nuclear degree of freedom for simplicity, the CBO ground-state
surface of \autoref{eq:VgsSC} may be written as:
\begin{equation}
V_0(R,q)=\frac{\omega_\nu^2}{2}R^2 + \frac{\omega_c^2}{2}q^2 + \lambda \omega_c q \mu(R),
\end{equation}
where we are here using mass-weighted coordinates ($R\to R/\sqrt{M}$) for the
molecular coordinate, which has a vibrational frequency of $\omega_\nu$. For
simplicity we are ignoring the polarizability term, which is equivalent to
redefining the effective photon frequency with the replacement $\frac12
\omega_c^2 (1-\lambda^2 \alpha(R)) q^2 \to \frac12 \omega_c^2 q^2$. The Hessian
of the surface is then
\begin{equation}
\mathcal{H} =
\begin{pmatrix}
\omega_\nu^2 & \lambda \omega_c \mu'_0(R_0)\\
\lambda \omega_c \mu'_0(R_0) & \omega_c^2
\end{pmatrix},
\end{equation}
where $\mu'_0(R_0)$ is the derivative of the ground-state dipole moment in
mass-weighted coordinates, evaluated at the minimum. The eigenvalues of the
Hessian correspond to the squares of the normal modes frequencies. In the
resonant case ($\omega_\nu = \omega_c$) it is straightforward to show that the
new frequencies are $\omega_\pm = \omega_c \sqrt{1\pm \frac{\lambda}{\omega_c}
\mu'_0(R_0)}$. This is the standard result for the modes of two coupled harmonic
oscillators beyond the rotating wave approximation~\cite{DelPino2015Signatures}.
The connection between the coupling strength and the Rabi splitting is clearer
in the limit of low coupling:
\begin{equation}
\omega_\pm \approx \omega_c \pm \frac{1}{2}\lambda \mu'_0(R_0).
\end{equation}
The Rabi splitting to lowest order is then $\Omega_\mathrm{R}=\omega_+ -
\omega_- = \lambda \mu'_0(R_0)$, i.e., proportional to $\lambda$, which is the
definition that we use in the main text. The Rabi splitting for collective
strong coupling is enhanced as $\Omega_\mathrm{R}= \sqrt{N} \lambda \mu'_0(R_0)$.

With the results above, it is trivial to obtain the zero-point energy in the
coupled system, which is given by $(\omega_+ + \omega_-)/2$. To second order in
$\lambda$, we obtain (without assuming resonance)
\begin{equation}
	E_{zp} = \frac12 (\omega_c + \omega_v) - \frac{\omega_c \lambda^2 \mu_0'^2(R_0)}{4 \omega_v (\omega_c + \omega_v)}.
\end{equation}
We note that when the effective frequency $\omega_{\mathrm{eff}}$ instead of the
bare frequency $\omega_c$ is used for the cavity mode, the additional correction
$-\frac{\lambda^2}{4}\omega_c\alpha(R_0)$ corresponds to an additive term up to
second order.

\section{Nanoparticle van der Waals potential}\label{app:nanoparticle_CP_vdW}
We here show that for a general nanoparticle with a series of (bosonic) dipole
resonances characterized by (vectorial) transition dipoles $\bmu_k$ and
frequencies $\omega_k$, the perturbative energy shift due to coupling to a
molecule, \autoref{eq:deltaE_pert}, corresponds exactly to Debye and London
forces. In this case, the coupling operators $\blambda_k$ at the molecular
position $\br_m$ are determined by the static dipole-dipole interaction,
\begin{equation}
	\blambda_k = \sqrt{\frac{2}{\omega_k}} \left(\frac{3 (\bmu_k \cdot \br_m) \br_m}{r_m^5} - \frac{\bmu_k}{r_m^3}\right).
\end{equation}
For simplicity, we assume $\br_m$ to be along the $x$-axis, and all dipoles to
be oriented along $z$, leading to
\begin{equation}
	\lambda_k = \sqrt{\frac{2}{\omega_k}} \frac{\mu_k}{r_m^3}.
\end{equation}
Inserting this into the perturbative energy correction then gives
\begin{multline}
	\delta E(\bR) = -\sum_k \frac{\lambda_k^2}{2} \left( \mu_0^2(\bR)
	+ \frac{\omega_k}{2} \alpha_0(\bR) \right)\\
	= -\sum_k \frac{\mu_k^2 \mu_0^2(\bR)}{\omega_k r_m^6} - \sum_k \frac{\mu_k^2 \alpha_0(\bR)}{2 r_m^6}.
\end{multline}
Using \autoref{eq:quant_polarizability}, the sum in the first term can be
replaced with the zero-frequency polarizability of the nanoparticle,
$\alpha_n(0) = \sum_k \frac{2\mu_k^2}{\omega_k}$, giving 
\begin{equation}
	\delta E(\bR) = -\frac{\alpha_q(0) \mu_0^2(\bR)}{2r_m^6} - \sum_k \frac{\mu_k^2 \alpha_0(\bR)}{2 r_m^6}.
\end{equation}
where the first term corresponds exactly to the static energy of a dipole
$\mu_0$ at $\br_m$ interacting with a polarizable sphere at the origin, and the
second term corresponds to the London force~\cite{Craig1998}.

\section{Electrostatics of a nanoparticle-on-mirror cavity}\label{app:dipole_images}
In here we derive the electrostatic energy of a dipole $\bmu$ inside a plasmonic
nanocavity made up of a spherical metallic nanoparticle of radius $R$ separated
by a gap $\Delta$ from a planar metallic mirror. This can be achieved using the
method of image charges by considering a formally infinite series of images,
with each image in a component of the cavity inducing an image in the other. In
practice, this infinite converging series can be truncated after a finite number
of terms to obtain any desired degree of accuracy. Considering both a charge $q$
and a dipole $\bmu$ at position $\br$ relative to the center of a perfectly
conducting grounded sphere of radius $R$, the resulting images will be located
at $\br' = (R/r)^2 \br$ (where $r=|\br|$) and consist of a charge and dipole
given by
\begin{align}
	q' &= -\frac{R}{r}q + \frac{R}{r^3} \br \cdot \bmu, \\
	\bmu' &= \left( \frac{R}{r} \right)^3 \left[ \frac{2 \br \left( \br \cdot \bmu \right)}{r^2} - \bmu \right].
\end{align}
Here, it is important to take into account that the image of a dipole in a
sphere always consists of both a charge and a dipole. The corresponding
expressions for a plane can be obtained by simply taking $R\to\infty$ (and
moving the center of the sphere accordingly to keep the planar surface fixed).
The cavity-induced energy shift of the dipole is then given by $U=-\frac{1}{2}
\bE_\mathrm{ind} \cdot \bmu$, where $\bE_\mathrm{ind}$ is the total field
generated by all image dipoles and charges, and the factor $\frac12$ is due to
them being induced.

It is also interesting to note that since a dipole induces a nonzero image
charge, the total induced dipole moment of the sphere is not origin-independent.
In particular, the induced dipole moment (for $q=0$) relative to the sphere
center is $\bmu' + \br' q' = (R/r)^3 \left[ 3 \br (\br\cdot\bmu)/r^2 -
\bmu\right]$, which corresponds to the dipole moment obtained when treating the
nanoparticle as a polarizable point particle
(cf.~\appref{app:sphere_quantization}). Accordingly, in a multipole expansion
about the sphere center, higher-order multipoles are nonzero and neglecting them
corresponds to an approximation, while using the image dipoles and charges as
given above is exact.

\end{appendix}

\bibliography{references}

\end{document}